\documentclass[reprint,amsmath,amssymb,aps,pre]{revtex4-1}

\usepackage{soul} 
\usepackage{graphicx}
\usepackage{dcolumn}
\usepackage{bm}
\usepackage{comment}
\usepackage{xcolor}

\newcommand{\specialcell}[2][c]{\footnotesize \begin{tabular}[#1]{@{}c@{}}#2\end{tabular}}

\begin{document}

\preprint{APS/123-QED}

\title{Topological generalization of the rigid-nonrigid transition in soft-sphere and hard-sphere fluids}

\author{Tae Jun Yoon}
\affiliation{School of Chemical and Biological Engineering, Institute of Chemical Processes, Seoul National University, Seoul 08826, Republic of Korea}

\author{Emanuel A. Lazar}
\affiliation{Department of Mathematics, Bar-Ilan University, Ramat Gan 5290002, Israel}

\author{Min Young Ha}
\affiliation{School of Chemical and Biological Engineering, Institute of Chemical Processes, Seoul National University, Seoul 08826, Republic of Korea}

\author{Won Bo Lee}
\email{wblee@snu.ac.kr}
\affiliation{School of Chemical and Biological Engineering, Institute of Chemical Processes, Seoul National University, Seoul 08826, Republic of Korea} 

\author{Youn-Woo Lee}
\email{ywlee@snu.ac.kr}
\affiliation{School of Chemical and Biological Engineering, Institute of Chemical Processes, Seoul National University, Seoul 08826, Republic of Korea}
 
\date{\today}

\begin{abstract}
A fluid particle changes its dynamics from diffusive to oscillatory as the system density increases up to the melting density. Hence, the notion of the Frenkel line was introduced to demarcate the fluid region into rigid and nonrigid liquid subregions based on the collective particle dynamics. In this work, we apply a topological framework to locate the Frenkel lines of the soft-sphere and the hard-sphere models relying on the system configurations. The topological characteristics of the ideal gas and the maximally random jammed state are first analyzed, then the classification scheme designed in our earlier work is applied. The classification result shows that the fraction of solid-like atoms increases from zero to one in the rigid liquid region. The dependence of the solid-like fraction on the bulk density is understood based on the theory of fluid polyamorphism. The percolation behavior of solid-like clusters is described based on the fraction of solid-like molecules in an integrated manner. The crossover densities are obtained by examining the percolation of solid-like clusters. The resultant crossover densities of soft-sphere fluids converge to that of hard-sphere fluid. Hence, the topological method successfully highlights the generality of the Frenkel line.
\end{abstract}

\maketitle

\section{Introduction}
\label{sec:level1}
Hard-sphere and soft-sphere models have been widely used to simulate the dynamics and structure of the fluid phase. The hard-sphere model is one of the most extensively studied model in statistical physics. The soft-sphere model, in which the penetrability of a sphere depends on the slope of the repulsive wall, has also been used as a simple model. In this model, the pair potential is given by
\begin{equation}
	\phi(r)=
  \begin{cases}
  	\epsilon\left({\sigma}/{r}\right)^n{\qquad}&r\leq\sigma\\
    0{\qquad}&r\geq\sigma\\
  \end{cases}
\end{equation}
where $\phi(r)$ is the interatomic pair potential between two spheres of diameter $\sigma$ separated by a distance $r$, $\epsilon$ is the energy parameter, and $n$ is the repulsive exponent. In these repulsive systems, no first-order gas-liquid transition occurs because the attractive interaction is absent. Hence, it is believed that the hard-sphere and the soft-sphere systems would follow the same dynamics scheme in the whole fluid region~\cite{bacher2014explaining}.

Brazhkin et al.~questioned this continuous picture of dynamics~\cite{brazhkin2012two}. Based on the phonon theory~\cite{bolmatov2012phonon,bolmatov2015unified} and on experimental validations~\cite{bolmatov2015frenkel}, they proposed the notion of the Frenkel line. They argued that the particle dynamics of supercritical fluid, a state of matter beyond the gas-liquid critical point, changes from diffusive (gas-like) to oscillatory (solid-like) across the Frenkel line. They proposed to use the heat capacity criterion and the velocity autocorrelation function to locate the Frenkel line of supercritical fluids~\cite{brazhkin2013liquid}. They further applied these thermodynamic and dynamic criteria to more general classes of fluid models such as the slightly soft-sphere model and the hard-sphere model.

However, a series of recent works on the phonon theory revealed that these thermodynamic and dynamic criteria could not be directly extended to locate the Frenkel line of all fluid models. Bryk et al.~noted that the Frenkel lines of the soft-sphere fluids located from these thermodynamic criteria did not converge to the density where the anomalous behavior of the hard-sphere models was observed~\cite{bryk2017non}. The non-convergence of the Frenkel lines of the soft-sphere fluids to that of the hard-sphere fluid was critical since the transport properties of the soft-sphere fluids including self-diffusion, shear and bulk viscosity, and thermal conductivity coefficients converge to those of the hard-sphere fluid~\cite{borgelt1989convergence,heyes2005transport}.

The non-convergence of the Frenkel lines mainly originates from the quasi-crystalline approximation (QCA), which is the basis of the conventional thermodynamic and dynamic criteria. Khrapak et al.~examined the validity of the QCA for the soft-sphere potentials and reported that the QCA fails when the repulsive exponent $n$ of the soft-sphere potential is higher than 20~\cite{khrapak2017collective}. Hence, Brazhkin et al.~recently proposed to use the anomalous transport properties as criteria to locate the rigid-nonrigid transition density of the hard-sphere model~\cite{brazhkin2018liquid}. Yoon et al.~proposed a dynamic criterion based on the notion of the solidicity from the two-phase thermodynamic (2PT) model~\cite{yoon2018two}. They observed that the solidicity, which is defined as the ratio of the diffusivity of a system to the diffusivity of the hard-sphere system at the zero pressure limit, shows an inflection behavior near the dynamic crossover density. Since the solidicity does not depend on the QCA, they demonstrated that the Frenkel lines of the soft spheres located from the new criterion converge to that of the hard-sphere model~\cite{brazhkin2018liquid}. 

In contrast to the dynamics-based approaches, there have been only a few works that attempted to locate the Frenkel line based not on the collective particle dynamics but on the system configuration. Bolmatov et al.~proposed that the third maximum of the pair correlation function be related to the structural crossover across the Frenkel line~\cite{bolmatov2014structural}. Ghosh et al.~applied this observation to characterize the Frenkel line of the confined fluids~\cite{ghosh2018structural}. However, Bryk et al.~criticized the use of the pair correlation function as a geometrical evidence of the Frenkel line since the intensity of the third maximum was so small that it cannot be distinguished from either the thermal noise or numerical errors~\cite{bryk2017behavior}. Fomin et al.~discovered that the packing fraction of effective hard spheres reaches the percolation threshold near the Frenkel line~\cite{fomin2014dynamic}. Ryltsev et al.~proposed that the concentration of tetrahedral clusters is higher than the percolation threshold above the Frenkel line~\cite{ryltsev2013multistage}. Unfortunately, they could explain how the solid-like structure evolves in the rigid liquid region only approximately. 

In our last work, thus, we suggested a topological classification procedure to analyze the dynamic crossover of supercritical argon across the Frenkel line~\cite{yoon2018topological}. In this procedure, geometric details of the Voronoi cells are not used to classify a particle as either gas-like or solid-like. Instead, the topological information, or the local connectivity of an atom to its neighbors, was used to classify whether an atom resembles the ideal gas or the maximally random jammed (MRJ) state. Based on this procedure, we discovered that the fraction of the solid-like particles steeply increases over the rigid liquid region, which is enclosed by the Frenkel line and the freezing line. This result provided the physical meaning of the Frenkel line as a percolation transition line from the ideal gas to the MRJ state. 

In this work, we apply the designed method to analyze the rigid-nonrigid crossover of the soft-sphere and hard-sphere fluids. We first examine the topological characteristics of the ideal gas and the MRJ state, respectively. The topological classification method is then applied to locate the Frenkel lines of soft-sphere and hard-sphere models. The dependence of the solid-like fraction on the bulk density is explained from the scope of the fluid polyamorphism theory and the isomorph theory. We further represent the generality of the dynamic crossover in terms of the solid-like fraction. The rigid-nonrigid crossover densities of soft-sphere fluids converge to that of the hard-sphere model as the repulsive exponent increases. These results substantiate that the topological framework can be used to locate the Frenkel line of general types of potentials.

\section{Methods}
\subsection{Molecular Dynamics (MD) simulations}
We performed time-driven NVT simulations~\cite{plimpton1995fast} of fluids modeled with the repulsive $n-6$ potentials [Eqn.~(\ref{eqn:n-6})]
\begin{equation}
	\phi_{n-6}(r) = C_{n}\epsilon\left[\left(\frac{\sigma}{r}\right)^n-\left(\frac{\sigma}{r}\right)^6\right]
  \label{eqn:n-6}
\end{equation}
where $C_{n}$ is given as:
\begin{equation}
	C_{n}=\left(\frac{n}{n-6}\right)\left(\frac{n}{6}\right)^{\frac{6}{n-6}}
\end{equation}
Each system contained 2,000 particles. The repulsive exponent $n$ was chosen from $n=8 - 32$. The size parameter ($\sigma$) was 3.405 \AA, the same as that of argon modeled with the Lennard-Jones potential. Energy parameters ($\epsilon$) were determined so that the coefficients $C_{n}\epsilon$ become equal to $4\epsilon_{lj}$ where $\epsilon_{lj}/k_{B}$ is equal to $119.8 \mbox{K}$, the energy parameter of argon. The potentials were shifted and truncated at the cutoff radius $r_{cut}=(n/6)^{1/(n-6)}\sigma$ where the $n-6$ potential has its minimum. The simulation temperatures were selected as $T^{*}=k_{B}T/\epsilon_{lj}=6.642 - 92.989$. The timestep was varied from $0.5 fs$ to $2 fs$ depending on the repulsive exponent ($n$) and the simulation conditions. The systems were equilibrated for 100,000 steps. After the equilibration, they were run for additional 100,000 steps to obtain the configurations and the system pressures. We additionally performed MD simulations of 16,000 particles modeled with the repulsive $12-6$ (Weeks-Chandler-Andersen \cite{weeks1971role}, WCA) potential at $T^{*}=92.989$ to examine the influence of the finite size effect. 

For the hard-sphere systems, the event-driven Molecular Dynamics (EDMD) simulations \cite{bannerman2011dynamo} were conducted. For each simulation condition, two hundred configurations of 2,048 hard spheres ($\sigma=1.0$, $k_{B}T=1.0$) were collected with a fixed dimensionless time interval of 1.00. The packing fractions ($\eta=\pi\rho\sigma^3/6$) were from 0.06 to 0.54 where $\rho$ is the number density of particles ($\rho=N/V$).

\subsection{Topological framework for the local structure analysis}
\begin{figure*}
	\includegraphics[width=\textwidth]{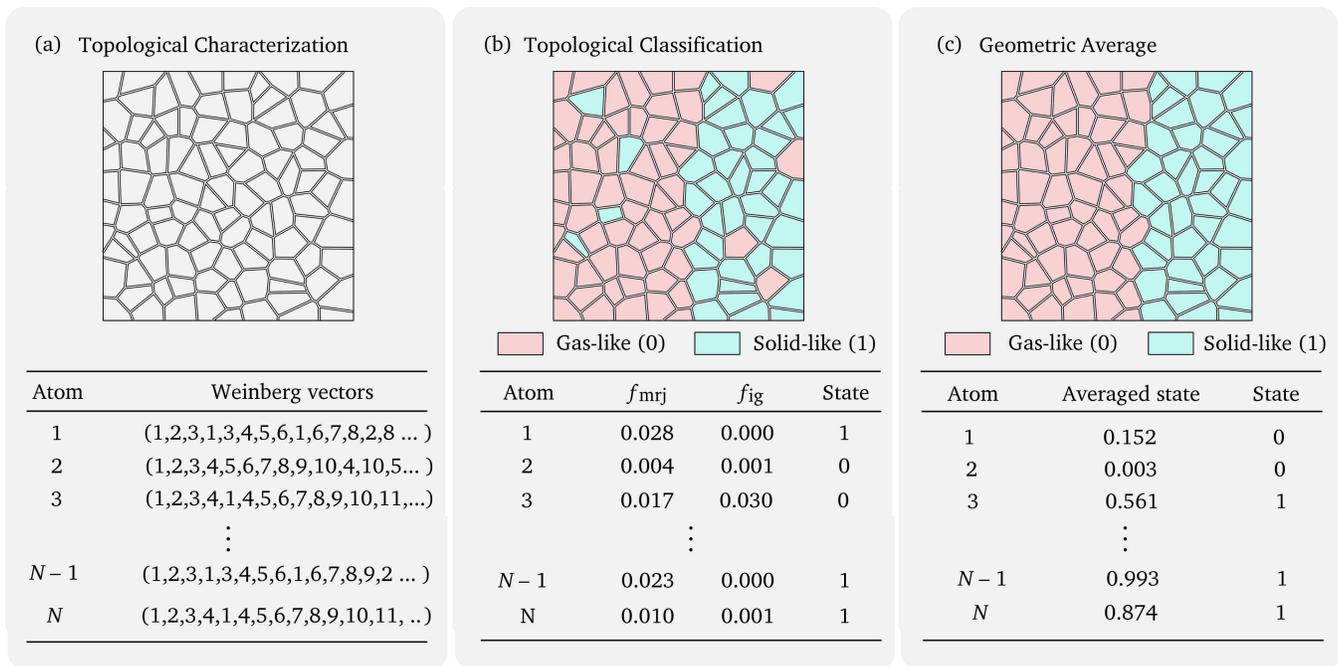}
  	\caption{A scheme of the topological classification procedure. In step (a), the types of Voronoi cells are obtained as a set of the Weinberg vectors. Then, the topological classification is performed based on the list of the Weinberg vectors of the ideal gas and the MRJ state. In step (b), a cell is classified based on the probabilities of finding the cell in the ideal gas and the MRJ state. After the topological classification, a weighted mean-field classification is conducted. The weight is given based on the chemical distance from the central atom. The Voronoi cells are represented as two-dimensional ones for understanding the algorithm conveniently.}
  \label{fig:top-geo-scheme}
\end{figure*}
The dynamics of a particle in a dense system is dominated by the relative positions of its surrounding neighbors. Hence, we conjectured that the topological framework for local structure analysis proposed by Lazar et al.~\cite{lazar2015topological} would build a robust connection between the dynamics and the geometry. In this framework, the arrangement of neighbors surrounding a central particle is described based on the Voronoi tessellation, the partitioning of space into regions which consist of all points closer to a central particle than to any other. A Voronoi cell provides the connectivity information as the geometric feature of a face and its adjacent faces. For instance, the number of faces of a Voronoi cell is equal to the number of nearest neighbors. An $n$-sided polygonal face of a Voronoi cell implies that the two atoms which share that face have $n$ common neighbors. The connectivity information of faces wrapping around a Voronoi cell can be encoded as a series of the integers called the Weinberg vector \cite{lazar2012complete}. Hence, the Weinberg vector can be viewed as a name of the type of a Voronoi cell. 

This topological description of the local configuration is in line with the isomorph theory. The isomorph theory \cite{schroder2014simplicity} states that the following simple relation is satisfied for any two configurations $\textbf{R}_{1}=(\vec{r_{1}}^{(1)},\cdots,\vec{r_{N}}^{(1)})$ and $\textbf{R}_{2}=(\vec{r_{1}}^{(2)},\cdots,\vec{r_{N}}^{(2)})$ if they are isomorphic to each other:
\begin{equation}
	\rho_{1}^{1/3}\textbf{R}_{1}=\rho_{2}^{1/3}\textbf{R}_{2} \implies P(\textbf{R}_{1})=P(\textbf{R}_{2})
\end{equation}
Here, $\vec{r}_{i}$ is the location vector of the particle $i$, $N$ is the number of atoms in the system, and $P$ is the Boltzmann statistical weight of the given configuration. In other words, two configurations are regarded to be isomorphic if their configurations in the reduced unit are the same. Similarly, the topological description does not depend on the distances between the central particle and its neighbors. Two local configurations are considered the same if their Voronoi cells are topologically identical regardless of their volumes.

\subsection{Topological characterization of the dynamic limits}
Since the Frenkel line was originally defined as the thermodynamic states where the particle dynamics change from diffusive to oscillatory, it was required to select the configurations that are opposite to each other from the viewpoint of the dynamics. A many-body system in which particles only translate without the interference of their neighbors would be the ideal gas. In contrast, systems in which most particles are randomly distributed and only vibrate in their place would be the MRJ state~\cite{klatt2014characterization}. Therefore, we constructed 50 samples of 500,000 particles each of both systems. The configurations of the ideal gas were generated by distributing points randomly, whereas those of the MRJ state were produced by compressing the hard-sphere system using the Lubachevsky-Stillinger algorithm \cite{lubachevsky1990geometric}. We then analyzed these samples using the open-source {\it VoroTop} software \cite{lazar2017vorotop}, which computed the distribution of topological features of the model systems.

\subsection{Topological classification strategy}
Fig.~\ref{fig:top-geo-scheme} schematically describes the topological classification strategy used in this work. The first step of the classification scheme is based on the list of the Weinberg vectors discovered in the dynamic limits. First, we compute the list of the Weinberg vectors of soft (hard) spheres obtained from the MD simulations using {\it VoroTop}. We then compare the likelihood of finding the Weinberg vector of a single atom in the list of the Weinberg vectors of the ideal gas ($f_{ig}$) to that of the MRJ state ($f_{mrj}$). If $f_{mrj} > f_{ig}$, the atom is classified as solid-like ($s_{i}=1$ where $s_{i}$ is a state number). Otherwise, it is classified as gas-like ($s_{i}=0$). After this initial classification, a weighted mean-field strategy \cite{yoon2018probabilistic} is used to reclassify an atom based on its state number and those of its neighbors. The averaged state number is defined as
\begin{equation}
	\bar{s}_{i}=\frac{1}{N_{i}}\sum_{j=0}^{N_{i}}\left(\frac{1}{N_{j}}\sum_{k=0}^{N_{j}}s_{k}\right)
\end{equation}
where $N_{i}$ is the number of the nearest neighbor atoms of the $i^{\rm{th}}$ particle. This procedure makes it possible to remove the influence of small local fluctuation on the classification result. Note that solid-like local structures might appear in ideal gas by pure chance, since the ideal gas essentially includes every possible configurations: if the topological framework alone is used, a nonzero fraction of particles in ideal gas are always classified as solid-like. By applying the weighted mean field procedure, the probability to find a solid-like molecule in the ideal gas converges to zero. As a last step, the averaged state number $\bar{s}_{i}$ is rounded to decide the state of molecule $i$. 

\subsection{Percolation analysis}
Percolation theory \cite{stauffer2014introduction} has frequently been used to analyze the structural characteristics of the connected molecules (clusters) discovered in the fluid phase. According to percolation theory, an infinite (spanning) cluster appears when the particle concentration is higher than a particular concentration of the particles called the percolation threshold ($\Pi^{c}$). Here, a cluster is defined as a connected assembly of atoms that are classified as the same state, either solid-like or gas-like. From the viewpoint of the Voronoi tessellation, two particles can be regarded to be connected if they share a face with each other. Hence, two atoms belong to the same cluster if they are connected through the atoms whose classification results are identical to theirs. A clustering algorithm proposed by Stoll \cite{stoll1998fast} is used to detect the cluster structure. In the first step of the algorithm, the list of clusters is obtained without consideration of the periodic boundary conditions. In the second step, the algorithm determines whether a cluster is infinite or not by examining the connectedness of two Voronoi cells which are located at opposite sides of the simulation box and assigned to the same cluster. If they are Voronoi neighbors across the boundary, the cluster is regarded to be a spanning (infinite) cluster. After this test, the independent clusters which are connected through the periodic boundaries are assigned to a single cluster if two atoms in these clusters are connected to each other.

\section{Results and Discussion}

\subsection{Topological characterization of the ideal gas and the MRJ state}
\begin{table}[b]
\centering
\makebox[0pt][c]{\parbox{0.95\columnwidth}{
\begin{minipage}[b]{0.45\hsize}\centering
\begin{tabular}{ | c | l | c |}
\hline
\multicolumn{3}{| c | }{{\bf Ideal gas }}\\ 
\hline
$F$ & {\bf $p$-vector} & $f(\%)$ \\ 
\hline
12 & $(1,3,4,3,1,0,\ldots)$ & 0.39 \\
11 & $(1,3,4,2,1,0,\ldots)$ & 0.34 \\
13 & $(1,4,3,3,2,0,\ldots)$ & 0.30 \\
13 & $(1,3,4,4,1,0,\ldots)$ & 0.29 \\
11 & $(1,4,2,3,1,0,\ldots)$ & 0.29 \\
13 & $(2,3,3,3,1,1,\ldots)$ & 0.28 \\
9 & $(1,3,3,2,0,0,\ldots)$ & 0.27 \\
10 & $(0,4,4,2,0,0,\ldots)$ & 0.26 \\
13 & $(1,3,5,2,2,0,\ldots)$ & 0.26 \\
11 & $(2,2,3,3,1,0,\ldots)$ & 0.26 \\
\hline    
\end{tabular}
\end{minipage}
\hfill
\begin{minipage}[b]{0.45\hsize}\centering
\begin{tabular}{ | c | l | c |}
\hline
\multicolumn{3}{| c | }{{\bf MRJ}}\\ \hline
$F$ & {\bf $p$-vector} & $f(\%)$ \\ \hline
13 & (0,3,6,4,0,\ldots) & 5.59 \\
14 & (0,2,8,4,0,\ldots) & 5.01 \\
14 & (0,3,6,5,0,\ldots) & 4.56 \\
13 & (0,1,10,2,0,\ldots) & 3.49 \\
14 & (0,4,4,6,0,\ldots) & 3.39 \\
14 & (1,3,4,5,1,\ldots) & 2.68 \\
15 & (0,3,6,6,0,\ldots) & 2.48 \\
15 & (0,2,8,5,0,\ldots) & 1.97 \\
13 & (0,2,8,3,0,\ldots) & 1.95 \\
12 & (0,2,8,2,0,\ldots) & 1.85 \\
\hline
\end{tabular}
\end{minipage}
}}
\caption{Lists of the ten most common $p$-vectors, their number of faces $F$, and their frequencies \emph{f} in the ideal gas and MRJ states.}
\label{table:pvectors}
\end{table}

\begin{table}
\centering
\makebox[0pt][c]{\parbox{1.\columnwidth}{
\begin{minipage}[b]{0.48\hsize}\centering
\begin{tabular}{ | l | r | r |}
\hline
$S$ & Ideal gas & MRJ \\ 
\hline
2	&	1,651,843	&	5,708,224	\\
3	&	1,841	&	75,323	\\
4	&	250,276	&	2,105,215	\\
6	&	70,600	&	740,474	\\
8	&	41,639	&	13,554	\\
10	&	89	&	17	\\
12	&	28,051	&	56,563	\\
16	&	7,872	&	467	\\
\hline    
\end{tabular}
\end{minipage}
\hfill
\begin{minipage}[b]{0.48\hsize}\centering
\begin{tabular}{ | l | r | r |}
\hline
$S$ & Ideal gas & MRJ \\ 
\hline
20	&	9,068	&	241	\\
24	&	7,281	&	55,375	\\
28	&	1,335	&	0	\\
32	&	273	&	0	\\
36	&	40	&	0	\\
40	&	2	&	0	\\
48	&	2,406	&	28,288	\\
120	&	383	&	335,454	\\
\hline
\end{tabular}
\end{minipage}
}}
\caption{Number of samples in each data set with symmetry order $S$. The symmetry order of the regular pentagonal dodecahedron (with icosohedral symmetry) is 120.}
\label{table:symmetries}
\end{table}

\setlength{\tabcolsep}{3pt}
\begin{figure*}
\centering
\begin{tabular}{|c|c|c|c|c|c|c|c|}
\hline
\multicolumn{8}{| c | }{{\bf Ideal gas }}\\ 
\hline
\specialcell[t]{1. $f$=0.27\%\\\includegraphics[scale=0.27]{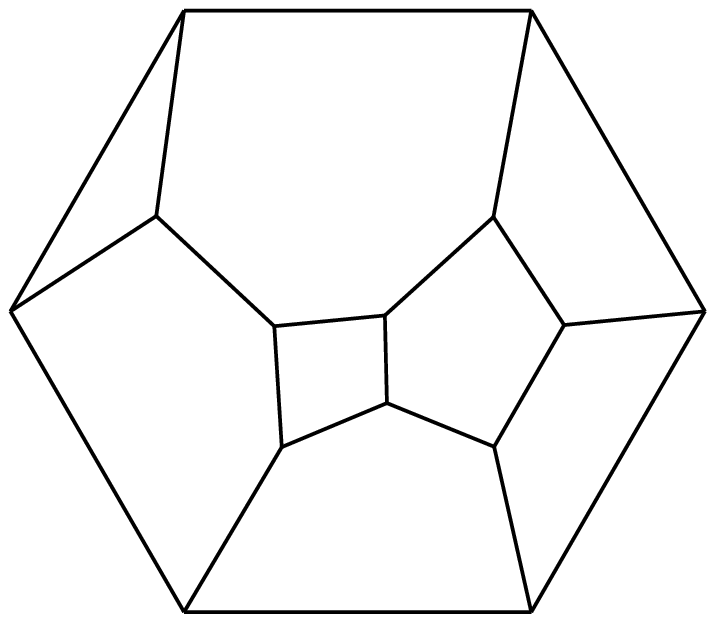}\\$(1,3,3,2,0,\ldots)$\\$F$=9, $S$=1 } & 
\specialcell[t]{2. $f$=0.16\%\\\includegraphics[scale=0.27]{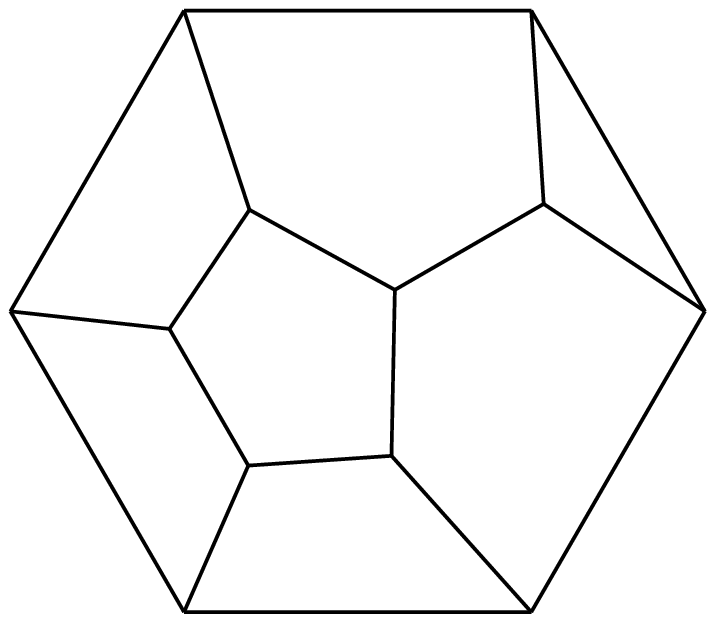}\\$(1,3,3,1,0,\ldots)$\\$F$=8, $S$=2 } & 
\specialcell[t]{3. $f$=0.16\%\\\includegraphics[scale=0.27]{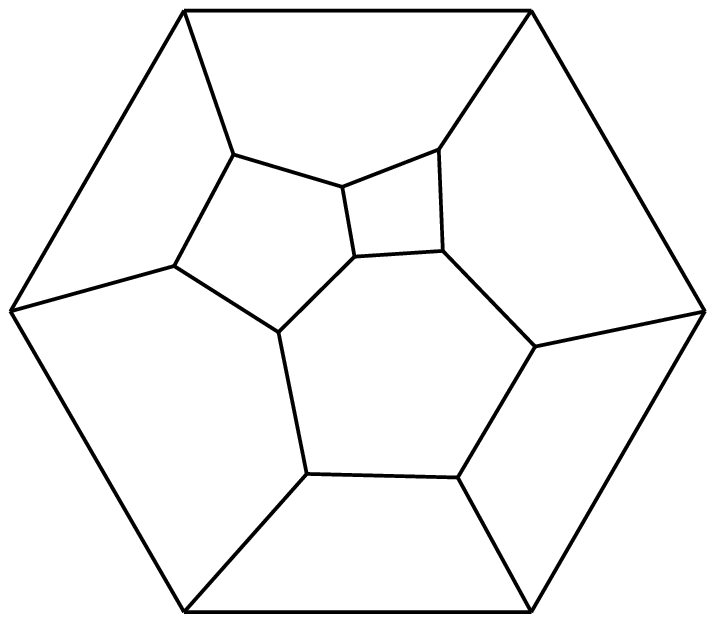}\\$(0,4,4,2,0,\ldots)$\\$F$=10, $S$=2 } & 
\specialcell[t]{4. $f$=0.12\%\\\includegraphics[scale=0.27]{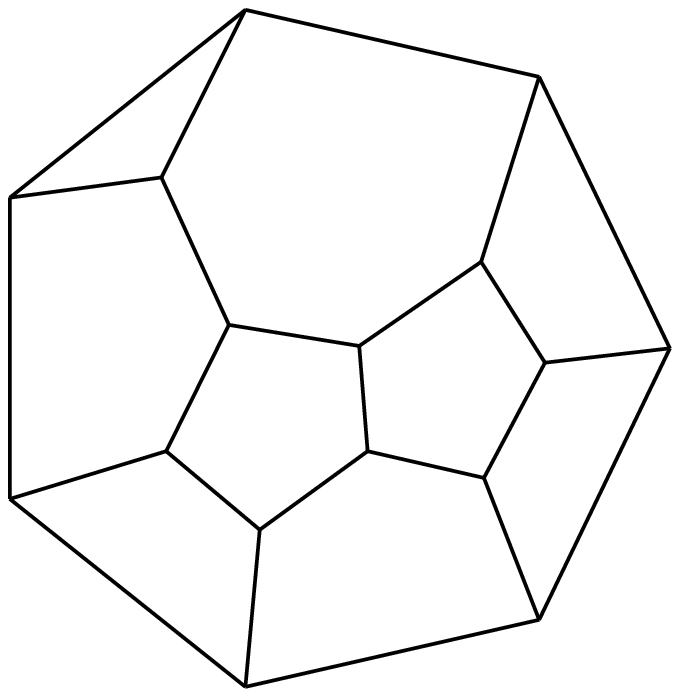}\\$(1,3,4,1,1,\ldots)$\\$F$=10, $S$=1 } & 
\specialcell[t]{5. $f$=0.12\%\\\includegraphics[scale=0.27]{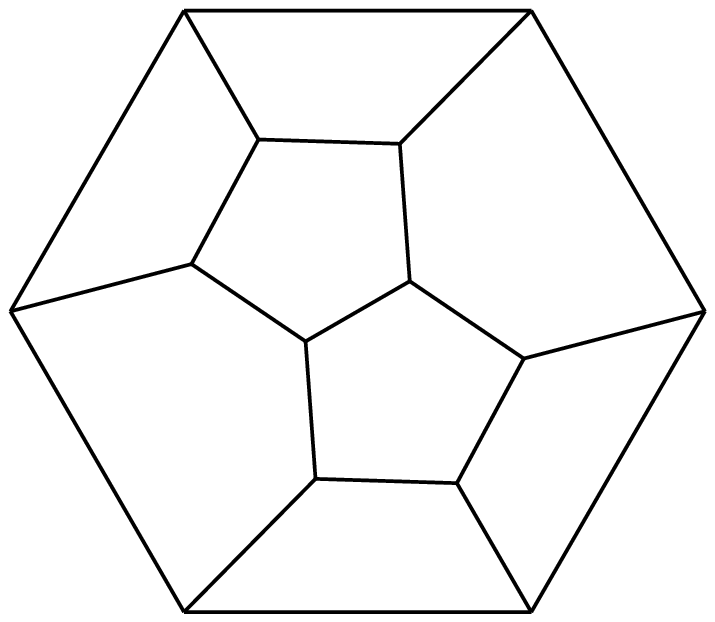}\\$(0,4,4,1,0,\ldots)$\\$F$=9, $S$=4 } & 
\specialcell[t]{6. $f$=0.10\%\\\includegraphics[scale=0.27]{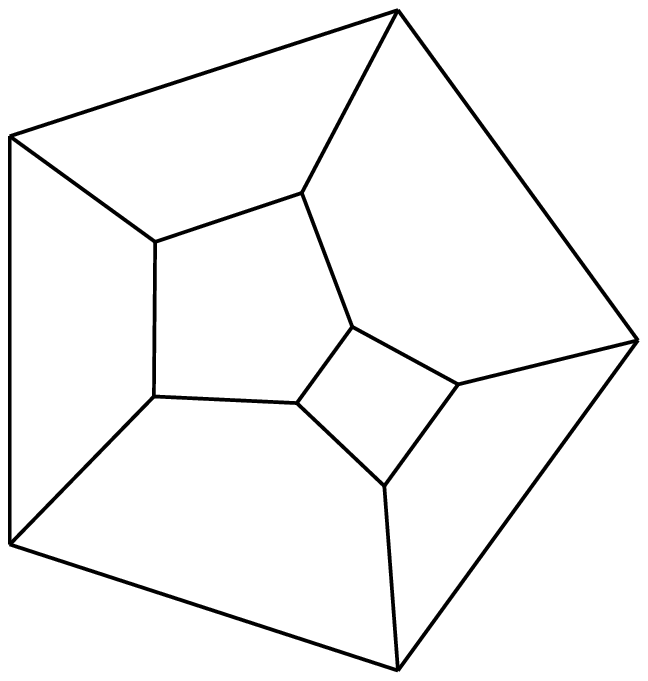}\\$(0,4,4,0,0,\ldots)$\\$F$=8, $S$=8 } & 
\specialcell[t]{7. $f$=0.10\%\\\includegraphics[scale=0.27]{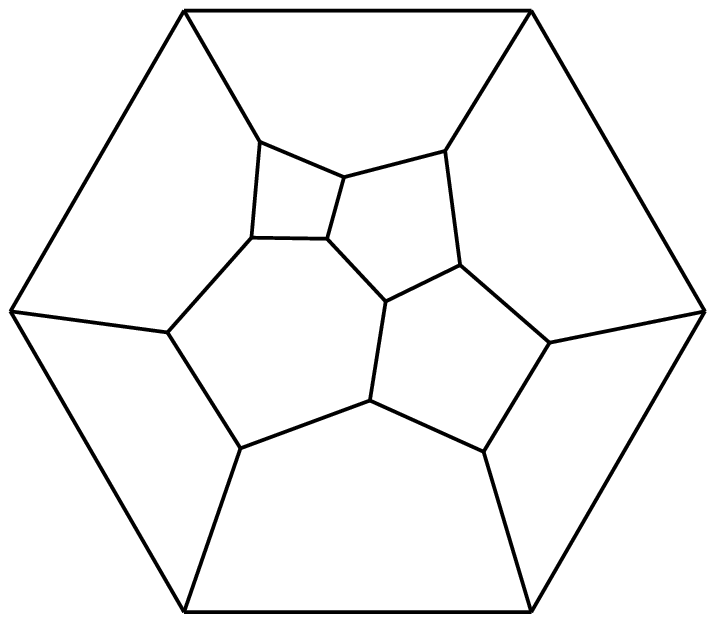}\\$(0,3,6,2,0,\ldots)$\\$F$=11, $S$=2 } & 
\specialcell[t]{8. $f$=0.10\%\\\includegraphics[scale=0.27]{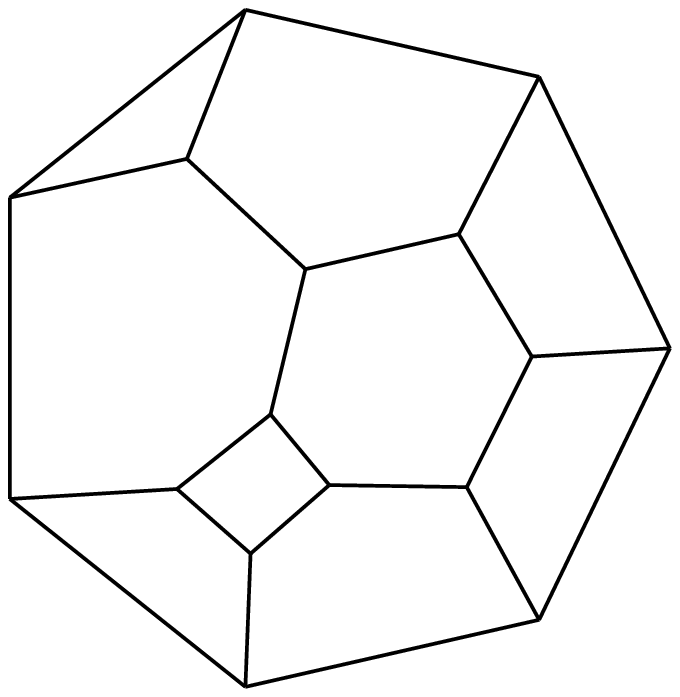}\\$(1,4,2,2,1,\ldots)$\\$F$=10, $S$=1 } \\
\hline
\hline
\multicolumn{8}{| c | }{{\bf MRJ }}\\ 
\hline
\specialcell[t]{1. $f$=3.49\%\\\includegraphics[scale=0.275]{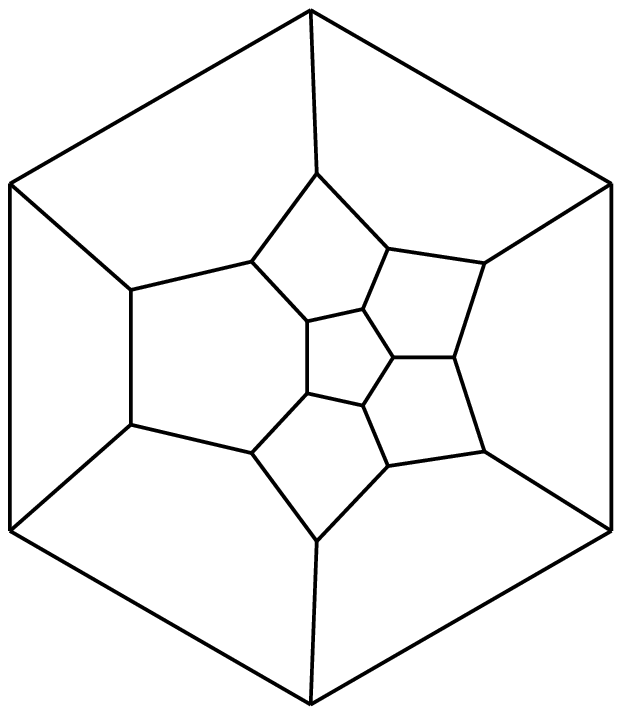}\\(0,1,10,2,\ldots)\\$F$=13, $S$=4 } & 
\specialcell[t]{2. $f$=2.40\%\\\includegraphics[scale=0.275]{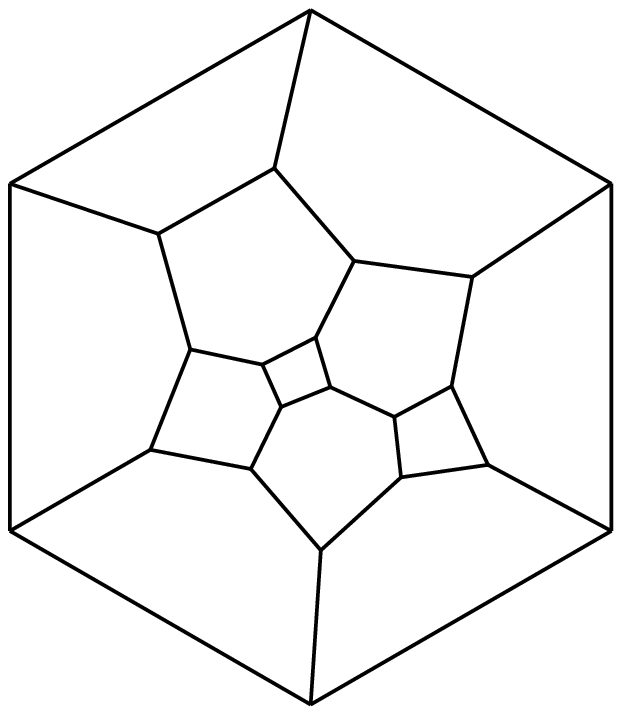}\\(0,3,6,4,\ldots)\\$F$=13, $S$=1 } & 
\specialcell[t]{3. $f$=2.06\%\\\includegraphics[scale=0.275]{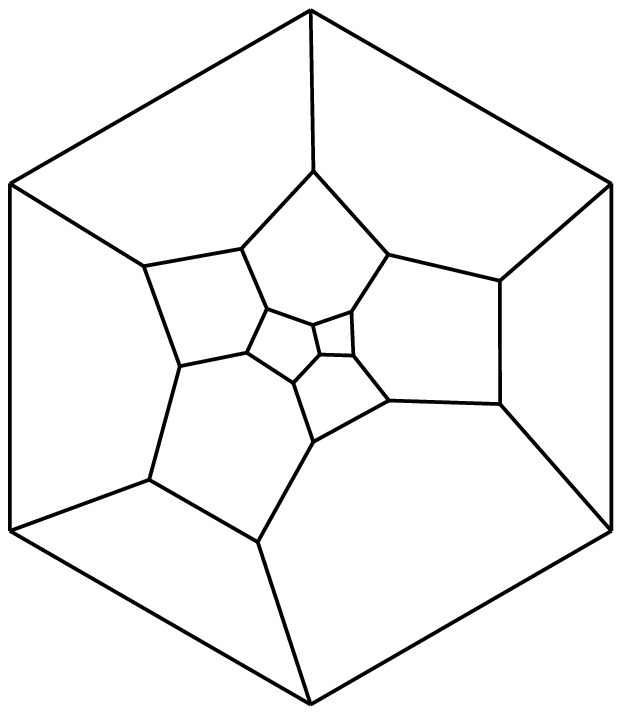}\\(0,3,6,5,\ldots)\\$F$=14, $S$=1 } & 
\specialcell[t]{4. $f$=1.76\%\\\includegraphics[scale=0.275]{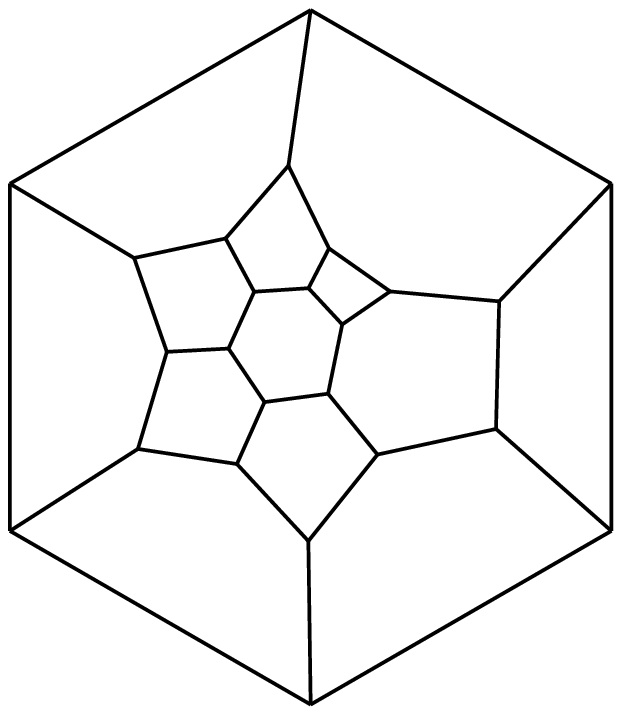}\\(0,2,8,4,\ldots)\\$F$=14, $S$=2 } & 
\specialcell[t]{5. $f$=1.53\%\\\includegraphics[scale=0.275]{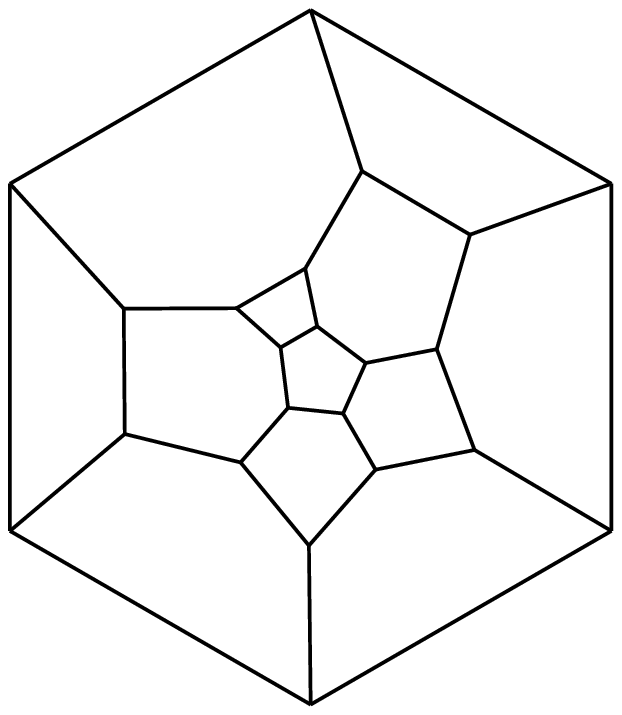}\\(0,3,6,4,\ldots)\\$F$=13, $S$=6 } & 
\specialcell[t]{6. $f$=1.48\%\\\includegraphics[scale=0.275]{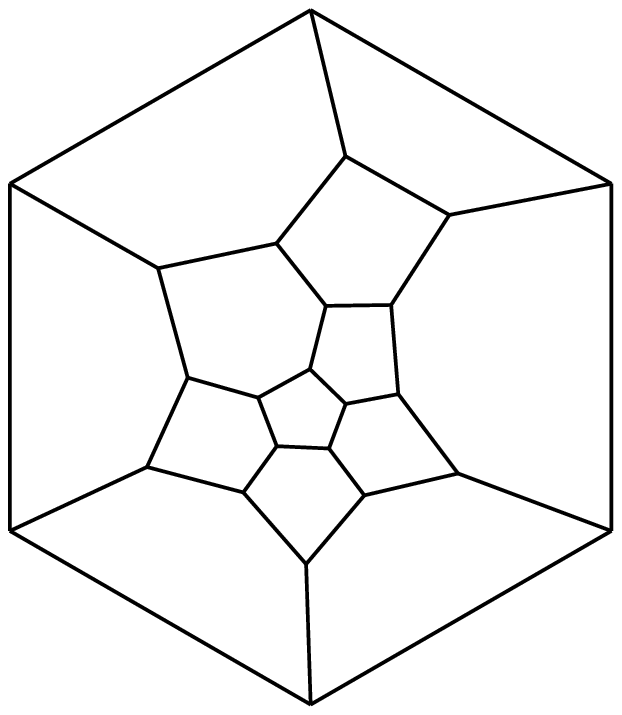}\\(0,1,10,3,\ldots)\\$F$=14, $S$=2 } & 
\specialcell[t]{7. $f$=1.48\%\\\includegraphics[scale=0.275]{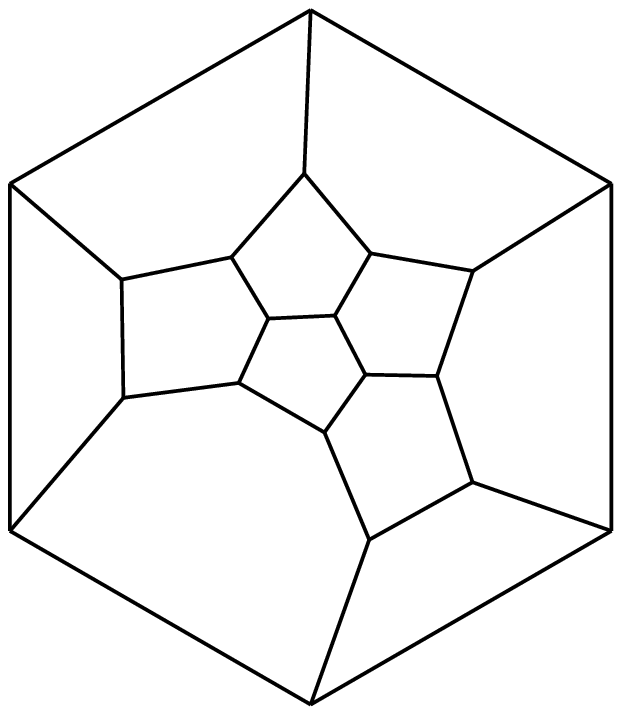}\\(0,2,8,2,\ldots)\\$F$=12, $S$=4 } & 
\specialcell[t]{8. $f$=1.44\%\\\includegraphics[scale=0.275]{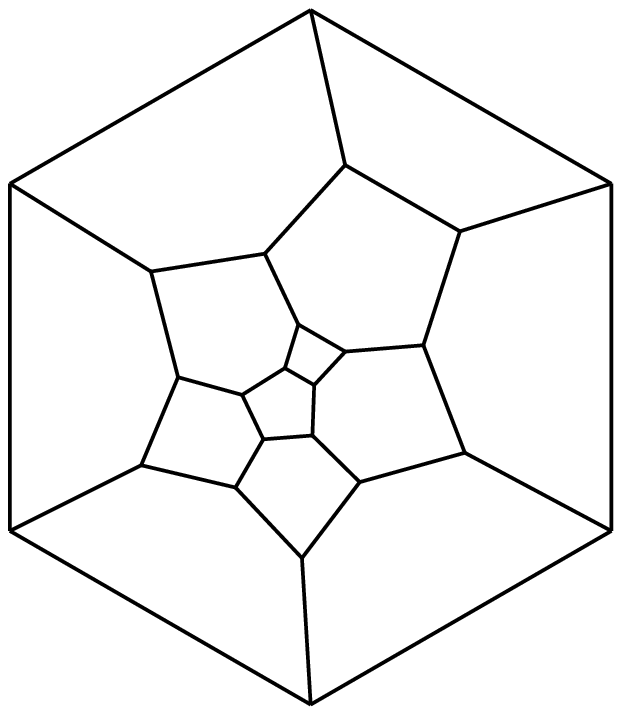}\\(0,2,8,4,\ldots)\\$F$=14, $S$=2 } \\
\hline
\end{tabular}
\caption{Schlegel diagrams of the 8 most common grain topologies (Weinberg vectors) in the ideal gas and MRJ systems. Listed for each topological type is the observed frequency $f$, the $p$-vector, the number of faces $F$, and the order $S$ of the associated symmetry group.}
\label{fig:schegels}
\end{figure*}

As the topology of each Voronoi cell encodes the manner in which neighbors of a particle are arranged \cite{lazar2015topological}, the distribution of Voronoi topologies provides a meaningful statistical description of local ordering in a system. In crystals, the set of possible Voronoi types is finite, restricted by the symmetry of the crystal and by the manner in which unstable vertices of an ideal structure can resolve \cite{lazar2015topological}. In contrast, the set of possible Voronoi types in an ideal gas is infinite, as almost all arrangements of particles are possible with some finite probability. Because the number of topological types is infinite, it is clear that not all arrangements are equally likely. Indeed, in prior work we documented the distribution of Voronoi topologies in the ideal gas and found that certain ones appear more frequently than others \cite{lazar2013statistical}. In this section we report data from the topological analysis of the ideal gas and MRJ systems.



The average number of faces in the ideal gas and MRJ systems were 15.54 ($\pm 3.33$) and 14.28 ($\pm 1.17$), respectively. The higher concentration of the distribution about the mean in the MRJ system suggests more order in that system than in the ideal gas; this similarly expresses itself in terms of the Voronoi indices and Voronoi topologies. Table \ref{table:pvectors} shows the 10 most common sets of Voronoi indices in the two system. Each set of indices, also called a $p$-vector \cite{lazar2012complete}, records the number of faces with a given number of edges, beginning with 3. The ten most frequent $p$-vectors in the MRJ system account for almost a third of all constituent atoms; in contrast, the ten most common $p$-vectors in the ideal gas account for less than three percent. The relative dispersion of $p$-vectors in the ideal gas case can be seen as reflecting the relative lack of order in that system, as compared with the MRJ one.

Fig. \ref{fig:schegels} illustrates the 8 most common Voronoi topologies in both the ideal gas and the MRJ systems, along with their estimated frequencies. The distribution of Voronoi topologies in the MRJ system is more concentrated among a smaller number of types, with each of the eight most frequent types accounting for at least 1.4\% of all atoms. In contrast, in the ideal gas, no single Voronoi topology accounts for more than 0.27\%. This can again be understood as reflecting the relative disorder of the ideal gas as compared with the MRJ system. 

Of additional potential interest in studying the relationship between the local neighborhood of a particle and its dynamics is the symmetry group associated with its Voronoi cell \cite{1966weinberg2}. 
The average symmetry group order is 1.161 in the ideal gas, and 3.365 in the MRJ system. Table \ref{table:symmetries} lists the number of atoms in each data set with non-trivial symmetry groups. The MRJ system has many more Voronoi cells with high-order symmetries than does the ideal gas. In particular, the number of atoms with icosahedral symmetry ($S=120$) is three orders of magnitude larger in the MRJ than in the ideal gas. 

The distribution of symmetry orders in the two systems might be correlated with their dynamics. In particular, the more symmetric a particle's Voronoi cell, the more confined that particle will be due to the influence of its symmetrically-arranged neighbors. For reference, in body-centered cubic crystals, the order of the symmetry group of each Voronoi cell is 48; other crystals have similarly high-order symmetries. Atoms with highly-symmetric arrangements of neighbors, and thus highly-symmetric Voronoi cells, might be thought of as trapped in a potential ``cage''.

Topological features of the Voronoi cells in the ideal gas and MRJ systems thus provide a quantitative description of local structure, and provide reference states against which to compare more general fluids. In a future paper we will provide a statistical description of a more general class of fluids.

\subsection{Frenkel lines of soft-sphere and hard-sphere fluids}
\begin{figure}
	\includegraphics{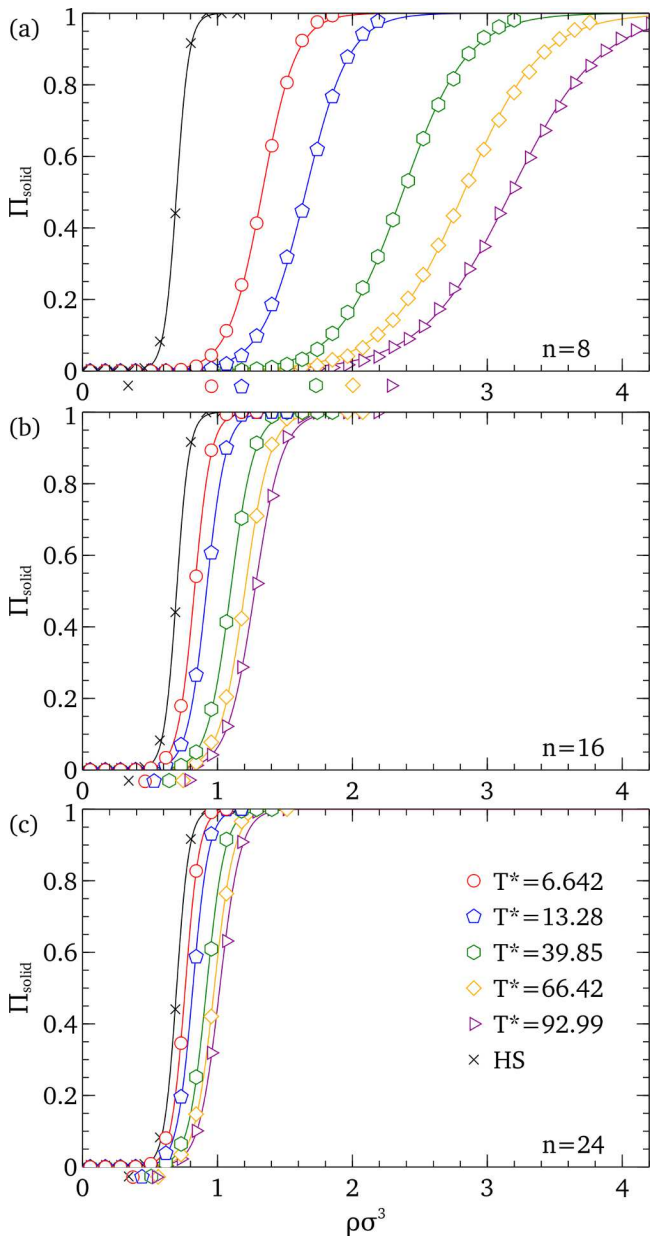}
  \caption{The fraction of solid-like molecules ($\Pi_{solid}$) as a function of the bulk density ($\rho$) at different conditions. As the repulsive exponent ($n$) increases, the temperature dependence of $\Pi_{solid}$ decreases. $\Pi_{solid}$ curves of the soft-sphere fluids become close to that of the hard-sphere fluid. The symbols below the horizontal axes denote the dynamic crossover densities obtained from the 2PT model.}
  \label{fig:fraction-solid}
\end{figure}

\begin{figure}
  \centering
  \includegraphics{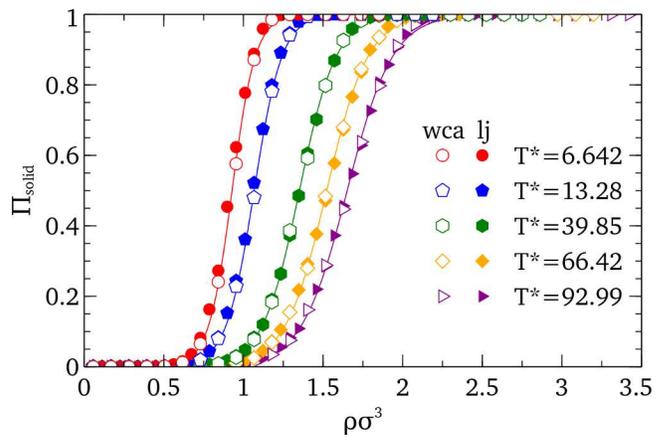}
  \caption{The fraction of solid-like molecules of the Lennard-Jones (LJ) and the repulsive 12-6 (WCA) fluids. The solid-like fraction curves of both systems agree with each other. The discrepancy of the solid-like fractions of both systems decreases as the temperature increases.}
  \label{fig:lj-comparison}
\end{figure}

Fig.~\ref{fig:fraction-solid} shows the fraction of solid-like molecules ($\Pi_{solid}$) of the soft-sphere fluids and the hard-sphere fluid. For all repulsive exponents and simulation temperatures, $\Pi_{solid}$'s show a sigmoidal dependence on the density [Eqn.~(\ref{eqn:Pi_solid_n})]. 
\begin{equation}
	\Pi_{solid}=\frac{1}{1+a\exp(-b\rho)}
  \label{eqn:Pi_solid_n}
\end{equation}
They start to steeply increase near the dynamic crossover densities obtained from the 2PT model and converge to unity near the freezing densities.

The sigmoidal dependence of $\Pi_{solid}$ on the density can be understood from the viewpoint of fluid polyamorphism \cite{anisimov2018thermodynamics,yoon2018probabilistic,ha2018widom}. According to the theory of fluid polyamorphism, the interconversion of gas-like and the solid-like states can be expressed as the chemical reaction ($A (gas) \rightleftharpoons A (solid)$). The equilibrium constant $K_{eq}$ of the interconversion is represented as
\begin{equation}
	K_{eq}=\frac{\Pi_{solid}}{\Pi_{gas}}=\exp\left(-\frac{{\Delta}G^\ddagger}{k_{B}T}\right)
  \label{eqn:K-eq}
\end{equation}
where ${\Delta}G^{\ddagger}$ is the Gibbs energy difference between the two states. Assuming that the Gibbs free energy of the interconversion (${\Delta}G^{\ddagger}$) is proportional to $\rho$, Eqn.~(\ref{eqn:K-eq}) can be transformed into Eqn.~(\ref{eqn:Pi_solid_n}). 

Figure~\ref{fig:fraction-solid} demonstrates that the structural evolution of soft-sphere fluids approaches that of the hard-sphere fluid as the repulsive exponent $n$ is increased, where the hard-sphere fluid is understood as the $n \rightarrow \infty$ limit. When the repulsive exponent $n$ is small (Fig.~\ref{fig:fraction-solid}a), $\Pi_{solid}$ curves largely depend on the temperatures. As the repulsive exponent increases (Fig.~\ref{fig:fraction-solid}b and c), different isothermal curves of $\Pi_{solid}$ more closely resemble that of the hard-sphere fluid. 

Figure~\ref{fig:fraction-solid} also shows that the crossover density is smaller in systems with larger $n$. As Brazhkin et al. stated \cite{brazhkin2018liquid}, the rigid-nonrigid crossover densities would be determined by the cage effect of surrounding molecules. The extent of the cage effect is determined by the competition between the relative kinetic energy of the central particle and the softness of the repulsive wall of its neighbors. When the repulsive exponent $n$ is high or the simulation temperature is low, particles are easily arrested by their nearest neighbors. Hence, the crossover density of the hard-sphere fluid should be lower than any soft-sphere fluids. When $n$ is small and the local cage is softer, more neighbors are required to trap the central particle at the high temperature. 

Interestingly, while the 2PT model and the theory of collective phonon dynamics based on the Frenkel frequency understand this cage effect based on the dynamics of a particle, the topological classification method quantifies it based on the fraction of solid-like molecules. This scenario implies that the attractive interaction would not play an important role in determining the location of the Frenkel line. Fig.~\ref{fig:lj-comparison} demonstrates this idea; the crossover densities of the LJ fluid from our earlier work, where the interatomic potential includes attraction, are almost consistent with those of the repulsive 12-6 fluid. When the system temperature is low, they slightly disagree with each other, but the discrepancy between these systems decreases as the temperature increases. Hence, the rigid-nonrigid crossover can be well explained by the hard-sphere paradigm, which states that the excluded volume effects dominate the liquid behavior \cite{weeks1971role,dyre2016simple}.
\begin{figure}
	\includegraphics{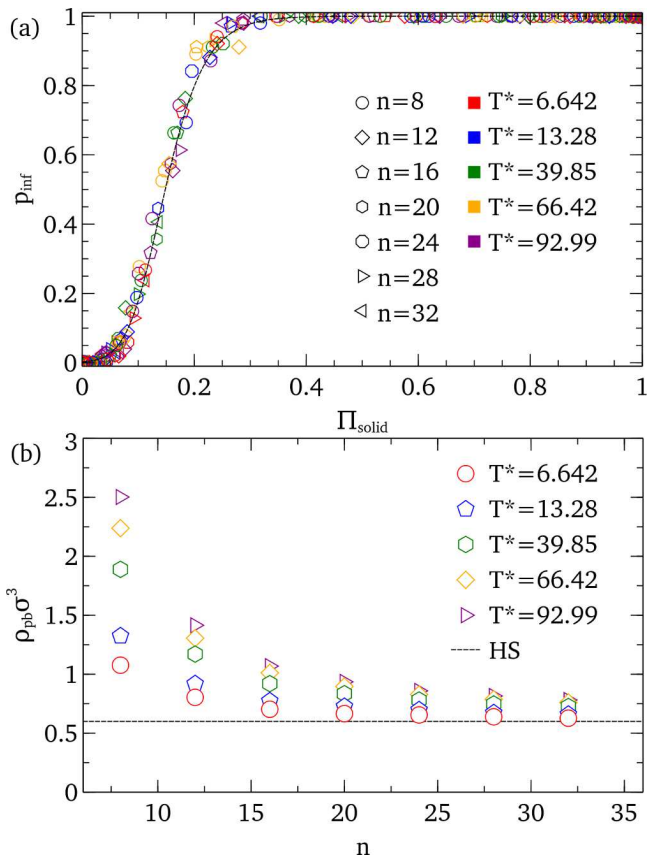}
  \caption{(a) The probability of finding an infinite cluster in a configuration ($p_{inf}$) as functions of the solid-like fraction ($\Pi_{solid}$). Regardless of the repulsive exponents and the temperatures, $p_{inf}$ collapses to a single line. The dotted line denotes the $p_{inf}$ curve obtained from our last work on the Lennard-Jones potential. (b) The convergence of the crossover densities (percolation densities) of the soft-sphere fluids to that of the hard-sphere fluid. As the repulsive exponent increases and the simulation temperature decreases, the crossover densities of soft spheres become close to that of hard spheres.}
  \label{fig:percolation}
\end{figure}

Next, we analyze the solid-like structures of the soft-sphere and the hard-sphere fluid systems from the viewpoint of the percolation theory. When the probability of finding an infinite cluster in a configuration ($p_{inf}$) is expressed as a function of the $\Pi_{solid}$, $p_{inf}$ curves at different conditions collapse to a single line obtained from the LJ simulations \cite{yoon2018topological} regardless of the repulsive exponents and temperatures (Fig.~\ref{fig:percolation}a). Moreover, both $p_{inf}$ curves of LJ and WCA fluid show the same dependence on the system size (see Fig. A1b in the Appendix).

This result demonstrates that the rigid-nonrigid transition across the Frenkel line is quasi-universal; it does not depend on the types of interactions. In line with the isomorph theory, it can be restated that two configurations with the same $\Pi_{solid}$ are isomorphic, showing the same percolation behavior. Provided that the percolation of solid-like clusters is related to the thermodynamic and transport properties of general fluid systems, the generality of the Frenkel line would have a deep relation with the excess entropy scaling \cite{rosenfeld1977relation,jakse2016excess}, which will be dealt with in our future studies. 

Since the percolation behaviors of the soft-sphere and hard-sphere fluids are equal to that of the LJ fluid, the percolation threshold obtained from our previous work ($\Pi_{solid}^{c}=0.1159\pm0.0081$) can be used to locate the dynamic crossover densities. Fig.~\ref{fig:percolation}b shows the crossover densities of soft-sphere and hard-sphere fluids. As shown in the 2PT method \cite{yoon2018two}, the dynamic crossover densities of soft-sphere fluids converge to that of hard-sphere fluid, which corresponds to $\eta=(\pi/6)\rho\sigma^3\sim0.315$. This result again demonstrates the advantage of the topological classification method. The thermodynamic and dynamic criteria by Brazhkin et al. are based on QCA, and the crossover densities from these criteria do not converge to that of the hard-sphere systems where QCA breaks down~\cite{bryk2017non}. The solidicity criterion from the 2PT model is free of QCA, yet it relies on Carnahan-Starling equation of state~\cite{carnahan1969equation}, which is an approximate model. On the other hand, the topological criterion from this work does not rely on hypotheses that are constrained to the repulsive exponents. Moreover, it does not require a vast amount of the post-processing procedure and data to obtain the thermodynamic properties of a system; it only requires the location of the particles and the simulation box length. Overall, these results show that the topological framework successfully generalizes the notion of the rigid-nonrigid transition in the fluid models.

\section{Conclusions}
The topological framework sheds light on the generalization of the notion of the rigid-nonrigid crossover. It not only offers physical insight into the relationship between the dynamics and the geometry of particles but also overcomes a limit of the conventional methods based on the thermodynamics and dynamics of the system. The dynamic limits of fluid particle systems can be clearly characterized based on their topological characteristics. The topological framework deduced from these dynamic limits provides a classification scheme that can locate the rigid-nonrigid transition of soft-sphere and hard-sphere systems. The fraction of solid-like molecules ($\Pi_{solid}$) from the classification method can be used as an order parameter to describe the rigid-nonrigid transition in an integrated manner. The dynamic crossover densities of the soft-sphere particles converge to that of the hard-sphere particles, which was also observed in the 2PT model. Hence, it would be advantageous to expand our understanding of the fluid physics as well as to calculate the thermodynamic properties of the fluid systems. 

\section*{Acknowledgements}
E. A. L. gratefully acknowledges the generous support of the US National Science Foundation through Award DMR-1507013. M.Y.H. acknowledges the support of the National Research Foundation of Korea Grants funded by Korean Government (NRF-2017H1A2A1044355). 

\section*{Appendix: Finite-size effect}
\renewcommand{\thefigure}{A\arabic{figure}}
\setcounter{figure}{0}
\begin{figure}
  \centering
  \includegraphics{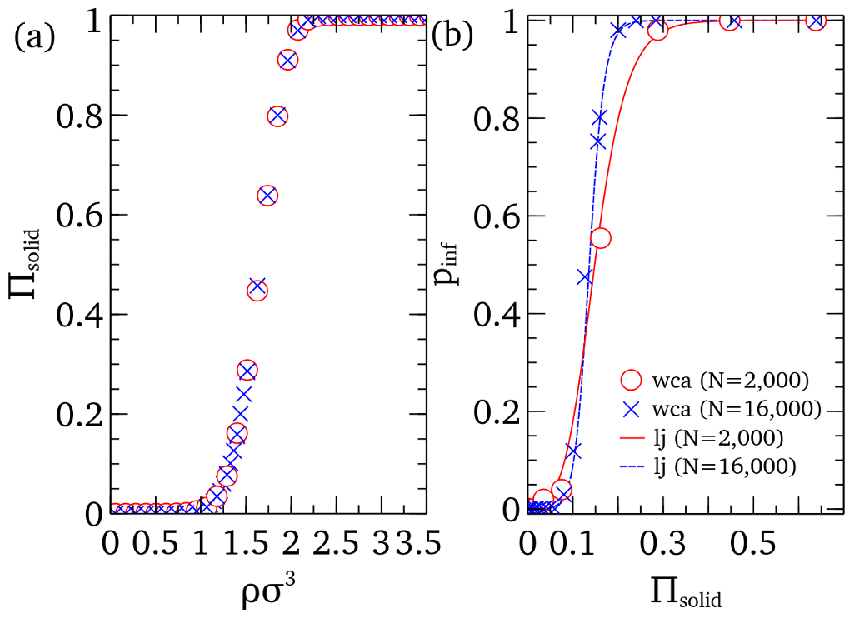}
  \caption{Finite-size effects on (a) the fraction of solid-like molecules and (b) the probability of finding an infinite cluster in a configuration. No significant finite-size effect on the solid-like fraction was observed. The percolation of solid-like structures, on the other hand, depends on the system size.}
  \label{fig:finite_size}
\end{figure}
Fig.~\ref{fig:finite_size}a compares the fraction of solid-like molecules modeled with the repulsive $12-6$ potential at $T^{*}=92.989$. The fraction of solid-like molecules does not change significantly when the number of molecules in the system increases. Hence, the topological classification results do not depend on the size of the system. In contrast, the percolation behavior of the system depends on the system size as shown in our earlier work \cite{yoon2018topological}. Compared to the system with 2,000 molecules, $p_{inf}$ of the $N=16,000$ system shows more abrupt increase when the system density increases. As shown in Fig.~\ref{fig:finite_size}b, the dependence of $p_{inf}$ on $\Pi_{solid}$ of the repulsive 12-6 (WCA) fluid is the same as that of the Lennard-Jones fluid. Again, this result substantiates that the percolation of solid-like structures in the rigid liquid region is universal.\\
\bibliography{bibfile}

\begin{thebibliography}{38}%
\makeatletter
\providecommand \@ifxundefined [1]{%
 \@ifx{#1\undefined}
}%
\providecommand \@ifnum [1]{%
 \ifnum #1\expandafter \@firstoftwo
 \else \expandafter \@secondoftwo
 \fi
}%
\providecommand \@ifx [1]{%
 \ifx #1\expandafter \@firstoftwo
 \else \expandafter \@secondoftwo
 \fi
}%
\providecommand \natexlab [1]{#1}%
\providecommand \enquote  [1]{``#1''}%
\providecommand \bibnamefont  [1]{#1}%
\providecommand \bibfnamefont [1]{#1}%
\providecommand \citenamefont [1]{#1}%
\providecommand \href@noop [0]{\@secondoftwo}%
\providecommand \href [0]{\begingroup \@sanitize@url \@href}%
\providecommand \@href[1]{\@@startlink{#1}\@@href}%
\providecommand \@@href[1]{\endgroup#1\@@endlink}%
\providecommand \@sanitize@url [0]{\catcode `\\12\catcode `\$12\catcode
  `\&12\catcode `\#12\catcode `\^12\catcode `\_12\catcode `\%12\relax}%
\providecommand \@@startlink[1]{}%
\providecommand \@@endlink[0]{}%
\providecommand \url  [0]{\begingroup\@sanitize@url \@url }%
\providecommand \@url [1]{\endgroup\@href {#1}{\urlprefix }}%
\providecommand \urlprefix  [0]{URL }%
\providecommand \Eprint [0]{\href }%
\providecommand \doibase [0]{http://dx.doi.org/}%
\providecommand \selectlanguage [0]{\@gobble}%
\providecommand \bibinfo  [0]{\@secondoftwo}%
\providecommand \bibfield  [0]{\@secondoftwo}%
\providecommand \translation [1]{[#1]}%
\providecommand \BibitemOpen [0]{}%
\providecommand \bibitemStop [0]{}%
\providecommand \bibitemNoStop [0]{.\EOS\space}%
\providecommand \EOS [0]{\spacefactor3000\relax}%
\providecommand \BibitemShut  [1]{\csname bibitem#1\endcsname}%
\let\auto@bib@innerbib\@empty
\bibitem [{\citenamefont {Bacher}\ \emph {et~al.}(2014)\citenamefont {Bacher},
  \citenamefont {Schr{\o}der},\ and\ \citenamefont
  {Dyre}}]{bacher2014explaining}%
  \BibitemOpen
  \bibfield  {author} {\bibinfo {author} {\bibfnamefont {A.~K.}\ \bibnamefont
  {Bacher}}, \bibinfo {author} {\bibfnamefont {T.~B.}\ \bibnamefont
  {Schr{\o}der}}, \ and\ \bibinfo {author} {\bibfnamefont {J.~C.}\ \bibnamefont
  {Dyre}},\ }\href@noop {} {\bibfield  {journal} {\bibinfo  {journal} {Nat.
  Commun.}\ }\textbf {\bibinfo {volume} {5}},\ \bibinfo {pages} {5424}
  (\bibinfo {year} {2014})}\BibitemShut {NoStop}%
\bibitem [{\citenamefont {Brazhkin}\ \emph {et~al.}(2012)\citenamefont
  {Brazhkin}, \citenamefont {Fomin}, \citenamefont {Lyapin}, \citenamefont
  {Ryzhov},\ and\ \citenamefont {Trachenko}}]{brazhkin2012two}%
  \BibitemOpen
  \bibfield  {author} {\bibinfo {author} {\bibfnamefont {V.}~\bibnamefont
  {Brazhkin}}, \bibinfo {author} {\bibfnamefont {Y.~D.}\ \bibnamefont {Fomin}},
  \bibinfo {author} {\bibfnamefont {A.}~\bibnamefont {Lyapin}}, \bibinfo
  {author} {\bibfnamefont {V.}~\bibnamefont {Ryzhov}}, \ and\ \bibinfo {author}
  {\bibfnamefont {K.}~\bibnamefont {Trachenko}},\ }\href@noop {} {\bibfield
  {journal} {\bibinfo  {journal} {Phys. Rev. E}\ }\textbf {\bibinfo {volume}
  {85}},\ \bibinfo {pages} {031203} (\bibinfo {year} {2012})}\BibitemShut
  {NoStop}%
\bibitem [{\citenamefont {Bolmatov}\ \emph {et~al.}(2012)\citenamefont
  {Bolmatov}, \citenamefont {Brazhkin},\ and\ \citenamefont
  {Trachenko}}]{bolmatov2012phonon}%
  \BibitemOpen
  \bibfield  {author} {\bibinfo {author} {\bibfnamefont {D.}~\bibnamefont
  {Bolmatov}}, \bibinfo {author} {\bibfnamefont {V.}~\bibnamefont {Brazhkin}},
  \ and\ \bibinfo {author} {\bibfnamefont {K.}~\bibnamefont {Trachenko}},\
  }\href@noop {} {\bibfield  {journal} {\bibinfo  {journal} {Sci. Rep.}\
  }\textbf {\bibinfo {volume} {2}},\ \bibinfo {pages} {421} (\bibinfo {year}
  {2012})}\BibitemShut {NoStop}%
\bibitem [{\citenamefont {Bolmatov}\ \emph
  {et~al.}(2015{\natexlab{a}})\citenamefont {Bolmatov}, \citenamefont
  {Zav’yalov}, \citenamefont {Zhernenkov}, \citenamefont {Musaev},\ and\
  \citenamefont {Cai}}]{bolmatov2015unified}%
  \BibitemOpen
  \bibfield  {author} {\bibinfo {author} {\bibfnamefont {D.}~\bibnamefont
  {Bolmatov}}, \bibinfo {author} {\bibfnamefont {D.}~\bibnamefont
  {Zav’yalov}}, \bibinfo {author} {\bibfnamefont {M.}~\bibnamefont
  {Zhernenkov}}, \bibinfo {author} {\bibfnamefont {E.~T.}\ \bibnamefont
  {Musaev}}, \ and\ \bibinfo {author} {\bibfnamefont {Y.~Q.}\ \bibnamefont
  {Cai}},\ }\href@noop {} {\bibfield  {journal} {\bibinfo  {journal} {Ann.
  Phys.}\ }\textbf {\bibinfo {volume} {363}},\ \bibinfo {pages} {221} (\bibinfo
  {year} {2015}{\natexlab{a}})}\BibitemShut {NoStop}%
\bibitem [{\citenamefont {Bolmatov}\ \emph
  {et~al.}(2015{\natexlab{b}})\citenamefont {Bolmatov}, \citenamefont
  {Zhernenkov}, \citenamefont {Zav’yalov}, \citenamefont {Tkachev},
  \citenamefont {Cunsolo},\ and\ \citenamefont {Cai}}]{bolmatov2015frenkel}%
  \BibitemOpen
  \bibfield  {author} {\bibinfo {author} {\bibfnamefont {D.}~\bibnamefont
  {Bolmatov}}, \bibinfo {author} {\bibfnamefont {M.}~\bibnamefont
  {Zhernenkov}}, \bibinfo {author} {\bibfnamefont {D.}~\bibnamefont
  {Zav’yalov}}, \bibinfo {author} {\bibfnamefont {S.~N.}\ \bibnamefont
  {Tkachev}}, \bibinfo {author} {\bibfnamefont {A.}~\bibnamefont {Cunsolo}}, \
  and\ \bibinfo {author} {\bibfnamefont {Y.~Q.}\ \bibnamefont {Cai}},\
  }\href@noop {} {\bibfield  {journal} {\bibinfo  {journal} {Sci. Rep.}\
  }\textbf {\bibinfo {volume} {5}},\ \bibinfo {pages} {15850} (\bibinfo {year}
  {2015}{\natexlab{b}})}\BibitemShut {NoStop}%
\bibitem [{\citenamefont {Brazhkin}\ \emph {et~al.}(2013)\citenamefont
  {Brazhkin}, \citenamefont {Fomin}, \citenamefont {Lyapin}, \citenamefont
  {Ryzhov}, \citenamefont {Tsiok},\ and\ \citenamefont
  {Trachenko}}]{brazhkin2013liquid}%
  \BibitemOpen
  \bibfield  {author} {\bibinfo {author} {\bibfnamefont {V.}~\bibnamefont
  {Brazhkin}}, \bibinfo {author} {\bibfnamefont {Y.~D.}\ \bibnamefont {Fomin}},
  \bibinfo {author} {\bibfnamefont {A.}~\bibnamefont {Lyapin}}, \bibinfo
  {author} {\bibfnamefont {V.}~\bibnamefont {Ryzhov}}, \bibinfo {author}
  {\bibfnamefont {E.}~\bibnamefont {Tsiok}}, \ and\ \bibinfo {author}
  {\bibfnamefont {K.}~\bibnamefont {Trachenko}},\ }\href@noop {} {\bibfield
  {journal} {\bibinfo  {journal} {Phys. Rev. Lett.}\ }\textbf {\bibinfo
  {volume} {111}},\ \bibinfo {pages} {145901} (\bibinfo {year}
  {2013})}\BibitemShut {NoStop}%
\bibitem [{\citenamefont {Bryk}\ \emph
  {et~al.}(2017{\natexlab{a}})\citenamefont {Bryk}, \citenamefont {Huerta},
  \citenamefont {Hordiichuk},\ and\ \citenamefont {Trokhymchuk}}]{bryk2017non}%
  \BibitemOpen
  \bibfield  {author} {\bibinfo {author} {\bibfnamefont {T.}~\bibnamefont
  {Bryk}}, \bibinfo {author} {\bibfnamefont {A.}~\bibnamefont {Huerta}},
  \bibinfo {author} {\bibfnamefont {V.}~\bibnamefont {Hordiichuk}}, \ and\
  \bibinfo {author} {\bibfnamefont {A.}~\bibnamefont {Trokhymchuk}},\
  }\href@noop {} {\bibfield  {journal} {\bibinfo  {journal} {J. Chem. Phys.}\
  }\textbf {\bibinfo {volume} {147}},\ \bibinfo {pages} {064509} (\bibinfo
  {year} {2017}{\natexlab{a}})}\BibitemShut {NoStop}%
\bibitem [{\citenamefont {Borgelt}\ and\ \citenamefont
  {Hoheisel}(1989)}]{borgelt1989convergence}%
  \BibitemOpen
  \bibfield  {author} {\bibinfo {author} {\bibfnamefont {P.}~\bibnamefont
  {Borgelt}}\ and\ \bibinfo {author} {\bibfnamefont {C.}~\bibnamefont
  {Hoheisel}},\ }\href@noop {} {\bibfield  {journal} {\bibinfo  {journal} {J.
  Chem. Phys.}\ }\textbf {\bibinfo {volume} {91}},\ \bibinfo {pages} {7872}
  (\bibinfo {year} {1989})}\BibitemShut {NoStop}%
\bibitem [{\citenamefont {Heyes}\ and\ \citenamefont
  {Bra{\'n}ka}(2005)}]{heyes2005transport}%
  \BibitemOpen
  \bibfield  {author} {\bibinfo {author} {\bibfnamefont {D.}~\bibnamefont
  {Heyes}}\ and\ \bibinfo {author} {\bibfnamefont {A.}~\bibnamefont
  {Bra{\'n}ka}},\ }\href@noop {} {\bibfield  {journal} {\bibinfo  {journal}
  {Phys. Chem. Chem. Phys.}\ }\textbf {\bibinfo {volume} {7}},\ \bibinfo
  {pages} {1220} (\bibinfo {year} {2005})}\BibitemShut {NoStop}%
\bibitem [{\citenamefont {Khrapak}\ \emph {et~al.}(2017)\citenamefont
  {Khrapak}, \citenamefont {Klumov},\ and\ \citenamefont
  {Cou{\"e}del}}]{khrapak2017collective}%
  \BibitemOpen
  \bibfield  {author} {\bibinfo {author} {\bibfnamefont {S.}~\bibnamefont
  {Khrapak}}, \bibinfo {author} {\bibfnamefont {B.}~\bibnamefont {Klumov}}, \
  and\ \bibinfo {author} {\bibfnamefont {L.}~\bibnamefont {Cou{\"e}del}},\
  }\href@noop {} {\bibfield  {journal} {\bibinfo  {journal} {Sci. Rep.}\
  }\textbf {\bibinfo {volume} {7}},\ \bibinfo {pages} {7985} (\bibinfo {year}
  {2017})}\BibitemShut {NoStop}%
\bibitem [{\citenamefont {Brazhkin}\ \emph {et~al.}(2018)\citenamefont
  {Brazhkin}, \citenamefont {Fomin}, \citenamefont {Ryzhov}, \citenamefont
  {Tsiok},\ and\ \citenamefont {Trachenko}}]{brazhkin2018liquid}%
  \BibitemOpen
  \bibfield  {author} {\bibinfo {author} {\bibfnamefont {V.}~\bibnamefont
  {Brazhkin}}, \bibinfo {author} {\bibfnamefont {Y.~D.}\ \bibnamefont {Fomin}},
  \bibinfo {author} {\bibfnamefont {V.}~\bibnamefont {Ryzhov}}, \bibinfo
  {author} {\bibfnamefont {E.}~\bibnamefont {Tsiok}}, \ and\ \bibinfo {author}
  {\bibfnamefont {K.}~\bibnamefont {Trachenko}},\ }\href@noop {} {\bibfield
  {journal} {\bibinfo  {journal} {Physica A}\ } (\bibinfo {year}
  {2018})}\BibitemShut {NoStop}%
\bibitem [{\citenamefont {Yoon}\ \emph
  {et~al.}(2018{\natexlab{a}})\citenamefont {Yoon}, \citenamefont {Ha},
  \citenamefont {Lee},\ and\ \citenamefont {Lee}}]{yoon2018two}%
  \BibitemOpen
  \bibfield  {author} {\bibinfo {author} {\bibfnamefont {T.~J.}\ \bibnamefont
  {Yoon}}, \bibinfo {author} {\bibfnamefont {M.~Y.}\ \bibnamefont {Ha}},
  \bibinfo {author} {\bibfnamefont {W.~B.}\ \bibnamefont {Lee}}, \ and\
  \bibinfo {author} {\bibfnamefont {Y.-W.}\ \bibnamefont {Lee}},\ }\href@noop
  {} {\bibfield  {journal} {\bibinfo  {journal} {arXiv preprint
  arXiv:1806.07608}\ } (\bibinfo {year} {2018}{\natexlab{a}})}\BibitemShut
  {NoStop}%
\bibitem [{\citenamefont {Bolmatov}\ \emph {et~al.}(2014)\citenamefont
  {Bolmatov}, \citenamefont {Zav’yalov}, \citenamefont {Gao},\ and\
  \citenamefont {Zhernenkov}}]{bolmatov2014structural}%
  \BibitemOpen
  \bibfield  {author} {\bibinfo {author} {\bibfnamefont {D.}~\bibnamefont
  {Bolmatov}}, \bibinfo {author} {\bibfnamefont {D.}~\bibnamefont
  {Zav’yalov}}, \bibinfo {author} {\bibfnamefont {M.}~\bibnamefont {Gao}}, \
  and\ \bibinfo {author} {\bibfnamefont {M.}~\bibnamefont {Zhernenkov}},\
  }\href@noop {} {\bibfield  {journal} {\bibinfo  {journal} {J. Phys. Chem.
  Lett.}\ }\textbf {\bibinfo {volume} {5}},\ \bibinfo {pages} {2785} (\bibinfo
  {year} {2014})}\BibitemShut {NoStop}%
\bibitem [{\citenamefont {Ghosh}\ and\ \citenamefont
  {Krishnamurthy}(2018)}]{ghosh2018structural}%
  \BibitemOpen
  \bibfield  {author} {\bibinfo {author} {\bibfnamefont {K.}~\bibnamefont
  {Ghosh}}\ and\ \bibinfo {author} {\bibfnamefont {C.}~\bibnamefont
  {Krishnamurthy}},\ }\href@noop {} {\bibfield  {journal} {\bibinfo  {journal}
  {Phys. Rev. E}\ }\textbf {\bibinfo {volume} {97}},\ \bibinfo {pages} {012131}
  (\bibinfo {year} {2018})}\BibitemShut {NoStop}%
\bibitem [{\citenamefont {Bryk}\ \emph
  {et~al.}(2017{\natexlab{b}})\citenamefont {Bryk}, \citenamefont {Gorelli},
  \citenamefont {Mryglod}, \citenamefont {Ruocco}, \citenamefont {Santoro},\
  and\ \citenamefont {Scopigno}}]{bryk2017behavior}%
  \BibitemOpen
  \bibfield  {author} {\bibinfo {author} {\bibfnamefont {T.}~\bibnamefont
  {Bryk}}, \bibinfo {author} {\bibfnamefont {F.~A.}\ \bibnamefont {Gorelli}},
  \bibinfo {author} {\bibfnamefont {I.}~\bibnamefont {Mryglod}}, \bibinfo
  {author} {\bibfnamefont {G.}~\bibnamefont {Ruocco}}, \bibinfo {author}
  {\bibfnamefont {M.}~\bibnamefont {Santoro}}, \ and\ \bibinfo {author}
  {\bibfnamefont {T.}~\bibnamefont {Scopigno}},\ }\href@noop {} {\bibfield
  {journal} {\bibinfo  {journal} {J. Phys. Chem. Lett.}\ }\textbf {\bibinfo
  {volume} {8}},\ \bibinfo {pages} {4995} (\bibinfo {year}
  {2017}{\natexlab{b}})}\BibitemShut {NoStop}%
\bibitem [{\citenamefont {Fomin}\ \emph {et~al.}(2014)\citenamefont {Fomin},
  \citenamefont {Ryzhov}, \citenamefont {Tsiok}, \citenamefont {Brazhkin},\
  and\ \citenamefont {Trachenko}}]{fomin2014dynamic}%
  \BibitemOpen
  \bibfield  {author} {\bibinfo {author} {\bibfnamefont {Y.~D.}\ \bibnamefont
  {Fomin}}, \bibinfo {author} {\bibfnamefont {V.}~\bibnamefont {Ryzhov}},
  \bibinfo {author} {\bibfnamefont {E.}~\bibnamefont {Tsiok}}, \bibinfo
  {author} {\bibfnamefont {V.}~\bibnamefont {Brazhkin}}, \ and\ \bibinfo
  {author} {\bibfnamefont {K.}~\bibnamefont {Trachenko}},\ }\href@noop {}
  {\bibfield  {journal} {\bibinfo  {journal} {Sci. Rep.}\ }\textbf {\bibinfo
  {volume} {4}},\ \bibinfo {pages} {7194} (\bibinfo {year} {2014})}\BibitemShut
  {NoStop}%
\bibitem [{\citenamefont {Ryltsev}\ and\ \citenamefont
  {Chtchelkatchev}(2013)}]{ryltsev2013multistage}%
  \BibitemOpen
  \bibfield  {author} {\bibinfo {author} {\bibfnamefont {R.}~\bibnamefont
  {Ryltsev}}\ and\ \bibinfo {author} {\bibfnamefont {N.}~\bibnamefont
  {Chtchelkatchev}},\ }\href@noop {} {\bibfield  {journal} {\bibinfo  {journal}
  {Phys. Rev. E}\ }\textbf {\bibinfo {volume} {88}},\ \bibinfo {pages} {052101}
  (\bibinfo {year} {2013})}\BibitemShut {NoStop}%
\bibitem [{\citenamefont {Yoon}\ \emph
  {et~al.}(2018{\natexlab{b}})\citenamefont {Yoon}, \citenamefont {Ha},
  \citenamefont {Lazar}, \citenamefont {Lee},\ and\ \citenamefont
  {Lee}}]{yoon2018topological}%
  \BibitemOpen
  \bibfield  {author} {\bibinfo {author} {\bibfnamefont {T.~J.}\ \bibnamefont
  {Yoon}}, \bibinfo {author} {\bibfnamefont {M.~Y.}\ \bibnamefont {Ha}},
  \bibinfo {author} {\bibfnamefont {E.~A.}\ \bibnamefont {Lazar}}, \bibinfo
  {author} {\bibfnamefont {W.~B.}\ \bibnamefont {Lee}}, \ and\ \bibinfo
  {author} {\bibfnamefont {Y.-W.}\ \bibnamefont {Lee}},\ }\href@noop {}
  {\bibfield  {journal} {\bibinfo  {journal} {arXiv preprint arXiv:1807.02761}\
  } (\bibinfo {year} {2018}{\natexlab{b}})}\BibitemShut {NoStop}%
\bibitem [{\citenamefont {Plimpton}(1995)}]{plimpton1995fast}%
  \BibitemOpen
  \bibfield  {author} {\bibinfo {author} {\bibfnamefont {S.}~\bibnamefont
  {Plimpton}},\ }\href@noop {} {\bibfield  {journal} {\bibinfo  {journal} {J.
  Comput. Phys.}\ }\textbf {\bibinfo {volume} {117}},\ \bibinfo {pages} {1}
  (\bibinfo {year} {1995})}\BibitemShut {NoStop}%
\bibitem [{\citenamefont {Weeks}\ \emph {et~al.}(1971)\citenamefont {Weeks},
  \citenamefont {Chandler},\ and\ \citenamefont {Andersen}}]{weeks1971role}%
  \BibitemOpen
  \bibfield  {author} {\bibinfo {author} {\bibfnamefont {J.~D.}\ \bibnamefont
  {Weeks}}, \bibinfo {author} {\bibfnamefont {D.}~\bibnamefont {Chandler}}, \
  and\ \bibinfo {author} {\bibfnamefont {H.~C.}\ \bibnamefont {Andersen}},\
  }\href@noop {} {\bibfield  {journal} {\bibinfo  {journal} {J. Chem. Phys.}\
  }\textbf {\bibinfo {volume} {54}},\ \bibinfo {pages} {5237} (\bibinfo {year}
  {1971})}\BibitemShut {NoStop}%
\bibitem [{\citenamefont {Bannerman}\ \emph {et~al.}(2011)\citenamefont
  {Bannerman}, \citenamefont {Sargant},\ and\ \citenamefont
  {Lue}}]{bannerman2011dynamo}%
  \BibitemOpen
  \bibfield  {author} {\bibinfo {author} {\bibfnamefont {M.~N.}\ \bibnamefont
  {Bannerman}}, \bibinfo {author} {\bibfnamefont {R.}~\bibnamefont {Sargant}},
  \ and\ \bibinfo {author} {\bibfnamefont {L.}~\bibnamefont {Lue}},\
  }\href@noop {} {\bibfield  {journal} {\bibinfo  {journal} {J. Comput. Chem.}\
  }\textbf {\bibinfo {volume} {32}},\ \bibinfo {pages} {3329} (\bibinfo {year}
  {2011})}\BibitemShut {NoStop}%
\bibitem [{\citenamefont {Lazar}\ \emph {et~al.}(2015)\citenamefont {Lazar},
  \citenamefont {Han},\ and\ \citenamefont {Srolovitz}}]{lazar2015topological}%
  \BibitemOpen
  \bibfield  {author} {\bibinfo {author} {\bibfnamefont {E.~A.}\ \bibnamefont
  {Lazar}}, \bibinfo {author} {\bibfnamefont {J.}~\bibnamefont {Han}}, \ and\
  \bibinfo {author} {\bibfnamefont {D.~J.}\ \bibnamefont {Srolovitz}},\
  }\href@noop {} {\bibfield  {journal} {\bibinfo  {journal} {Proc. Natl. Acad.
  Sci. U.S.A.}\ }\textbf {\bibinfo {volume} {112}},\ \bibinfo {pages} {E5769}
  (\bibinfo {year} {2015})}\BibitemShut {NoStop}%
\bibitem [{\citenamefont {Lazar}\ \emph {et~al.}(2012)\citenamefont {Lazar},
  \citenamefont {Mason}, \citenamefont {MacPherson},\ and\ \citenamefont
  {Srolovitz}}]{lazar2012complete}%
  \BibitemOpen
  \bibfield  {author} {\bibinfo {author} {\bibfnamefont {E.~A.}\ \bibnamefont
  {Lazar}}, \bibinfo {author} {\bibfnamefont {J.~K.}\ \bibnamefont {Mason}},
  \bibinfo {author} {\bibfnamefont {R.~D.}\ \bibnamefont {MacPherson}}, \ and\
  \bibinfo {author} {\bibfnamefont {D.~J.}\ \bibnamefont {Srolovitz}},\
  }\href@noop {} {\bibfield  {journal} {\bibinfo  {journal} {Phys. Rev. Lett.}\
  }\textbf {\bibinfo {volume} {109}},\ \bibinfo {pages} {095505} (\bibinfo
  {year} {2012})}\BibitemShut {NoStop}%
\bibitem [{\citenamefont {Schr{\o}der}\ and\ \citenamefont
  {Dyre}(2014)}]{schroder2014simplicity}%
  \BibitemOpen
  \bibfield  {author} {\bibinfo {author} {\bibfnamefont {T.~B.}\ \bibnamefont
  {Schr{\o}der}}\ and\ \bibinfo {author} {\bibfnamefont {J.~C.}\ \bibnamefont
  {Dyre}},\ }\href@noop {} {\bibfield  {journal} {\bibinfo  {journal} {J. Chem.
  Phys.}\ }\textbf {\bibinfo {volume} {141}},\ \bibinfo {pages} {204502}
  (\bibinfo {year} {2014})}\BibitemShut {NoStop}%
\bibitem [{\citenamefont {Klatt}\ and\ \citenamefont
  {Torquato}(2014)}]{klatt2014characterization}%
  \BibitemOpen
  \bibfield  {author} {\bibinfo {author} {\bibfnamefont {M.~A.}\ \bibnamefont
  {Klatt}}\ and\ \bibinfo {author} {\bibfnamefont {S.}~\bibnamefont
  {Torquato}},\ }\href@noop {} {\bibfield  {journal} {\bibinfo  {journal}
  {Phys. Rev. E}\ }\textbf {\bibinfo {volume} {90}},\ \bibinfo {pages} {052120}
  (\bibinfo {year} {2014})}\BibitemShut {NoStop}%
\bibitem [{\citenamefont {Lubachevsky}\ and\ \citenamefont
  {Stillinger}(1990)}]{lubachevsky1990geometric}%
  \BibitemOpen
  \bibfield  {author} {\bibinfo {author} {\bibfnamefont {B.~D.}\ \bibnamefont
  {Lubachevsky}}\ and\ \bibinfo {author} {\bibfnamefont {F.~H.}\ \bibnamefont
  {Stillinger}},\ }\href@noop {} {\bibfield  {journal} {\bibinfo  {journal} {J.
  Stat. Phys.}\ }\textbf {\bibinfo {volume} {60}},\ \bibinfo {pages} {561}
  (\bibinfo {year} {1990})}\BibitemShut {NoStop}%
\bibitem [{\citenamefont {Lazar}(2017)}]{lazar2017vorotop}%
  \BibitemOpen
  \bibfield  {author} {\bibinfo {author} {\bibfnamefont {E.~A.}\ \bibnamefont
  {Lazar}},\ }\href@noop {} {\bibfield  {journal} {\bibinfo  {journal}
  {Modelling Simul. Mater. Sci. Eng.}\ }\textbf {\bibinfo {volume} {26}},\
  \bibinfo {pages} {015011} (\bibinfo {year} {2017})}\BibitemShut {NoStop}%
\bibitem [{\citenamefont {Yoon}\ \emph
  {et~al.}(2018{\natexlab{c}})\citenamefont {Yoon}, \citenamefont {Ha},
  \citenamefont {Lee},\ and\ \citenamefont {Lee}}]{yoon2018probabilistic}%
  \BibitemOpen
  \bibfield  {author} {\bibinfo {author} {\bibfnamefont {T.~J.}\ \bibnamefont
  {Yoon}}, \bibinfo {author} {\bibfnamefont {M.~Y.}\ \bibnamefont {Ha}},
  \bibinfo {author} {\bibfnamefont {W.~B.}\ \bibnamefont {Lee}}, \ and\
  \bibinfo {author} {\bibfnamefont {Y.-W.}\ \bibnamefont {Lee}},\ }\href@noop
  {} {\bibfield  {journal} {\bibinfo  {journal} {J. Chem. Phys.}\ }\textbf
  {\bibinfo {volume} {149}},\ \bibinfo {pages} {014502} (\bibinfo {year}
  {2018}{\natexlab{c}})}\BibitemShut {NoStop}%
\bibitem [{\citenamefont {Stauffer}\ and\ \citenamefont
  {Aharony}(2014)}]{stauffer2014introduction}%
  \BibitemOpen
  \bibfield  {author} {\bibinfo {author} {\bibfnamefont {D.}~\bibnamefont
  {Stauffer}}\ and\ \bibinfo {author} {\bibfnamefont {A.}~\bibnamefont
  {Aharony}},\ }\href@noop {} {\emph {\bibinfo {title} {Introduction to
  Percolation Theory: Revised Second Edition}}}\ (\bibinfo  {publisher} {CRC
  press},\ \bibinfo {year} {2014})\BibitemShut {NoStop}%
\bibitem [{\citenamefont {Stoll}(1998)}]{stoll1998fast}%
  \BibitemOpen
  \bibfield  {author} {\bibinfo {author} {\bibfnamefont {E.}~\bibnamefont
  {Stoll}},\ }\href@noop {} {\bibfield  {journal} {\bibinfo  {journal} {Comput.
  Phys. Commun.}\ }\textbf {\bibinfo {volume} {109}},\ \bibinfo {pages} {1}
  (\bibinfo {year} {1998})}\BibitemShut {NoStop}%
\bibitem [{\citenamefont {Lazar}\ \emph {et~al.}(2013)\citenamefont {Lazar},
  \citenamefont {Mason}, \citenamefont {MacPherson},\ and\ \citenamefont
  {Srolovitz}}]{lazar2013statistical}%
  \BibitemOpen
  \bibfield  {author} {\bibinfo {author} {\bibfnamefont {E.~A.}\ \bibnamefont
  {Lazar}}, \bibinfo {author} {\bibfnamefont {J.~K.}\ \bibnamefont {Mason}},
  \bibinfo {author} {\bibfnamefont {R.~D.}\ \bibnamefont {MacPherson}}, \ and\
  \bibinfo {author} {\bibfnamefont {D.~J.}\ \bibnamefont {Srolovitz}},\
  }\href@noop {} {\bibfield  {journal} {\bibinfo  {journal} {Phys. Rev. E}\
  }\textbf {\bibinfo {volume} {88}},\ \bibinfo {pages} {063309} (\bibinfo
  {year} {2013})}\BibitemShut {NoStop}%
\bibitem [{\citenamefont {Weinberg}(1966)}]{1966weinberg2}%
  \BibitemOpen
  \bibfield  {author} {\bibinfo {author} {\bibfnamefont {L.}~\bibnamefont
  {Weinberg}},\ }\href@noop {} {\bibfield  {journal} {\bibinfo  {journal} {SIAM
  J. Appl. Math.}\ }\textbf {\bibinfo {volume} {{14}}},\ \bibinfo {pages} {729}
  (\bibinfo {year} {1966})}\BibitemShut {NoStop}%
\bibitem [{\citenamefont {Anisimov}\ \emph {et~al.}(2018)\citenamefont
  {Anisimov}, \citenamefont {Du{\v{s}}ka}, \citenamefont {Caupin},
  \citenamefont {Amrhein}, \citenamefont {Rosenbaum},\ and\ \citenamefont
  {Sadus}}]{anisimov2018thermodynamics}%
  \BibitemOpen
  \bibfield  {author} {\bibinfo {author} {\bibfnamefont {M.~A.}\ \bibnamefont
  {Anisimov}}, \bibinfo {author} {\bibfnamefont {M.}~\bibnamefont
  {Du{\v{s}}ka}}, \bibinfo {author} {\bibfnamefont {F.}~\bibnamefont {Caupin}},
  \bibinfo {author} {\bibfnamefont {L.~E.}\ \bibnamefont {Amrhein}}, \bibinfo
  {author} {\bibfnamefont {A.}~\bibnamefont {Rosenbaum}}, \ and\ \bibinfo
  {author} {\bibfnamefont {R.~J.}\ \bibnamefont {Sadus}},\ }\href@noop {}
  {\bibfield  {journal} {\bibinfo  {journal} {Phys. Rev. X}\ }\textbf {\bibinfo
  {volume} {8}},\ \bibinfo {pages} {011004} (\bibinfo {year}
  {2018})}\BibitemShut {NoStop}%
\bibitem [{\citenamefont {Ha}\ \emph {et~al.}(2018)\citenamefont {Ha},
  \citenamefont {Yoon}, \citenamefont {Tlusty}, \citenamefont {Jho},\ and\
  \citenamefont {Lee}}]{ha2018widom}%
  \BibitemOpen
  \bibfield  {author} {\bibinfo {author} {\bibfnamefont {M.~Y.}\ \bibnamefont
  {Ha}}, \bibinfo {author} {\bibfnamefont {T.~J.}\ \bibnamefont {Yoon}},
  \bibinfo {author} {\bibfnamefont {T.}~\bibnamefont {Tlusty}}, \bibinfo
  {author} {\bibfnamefont {Y.}~\bibnamefont {Jho}}, \ and\ \bibinfo {author}
  {\bibfnamefont {W.~B.}\ \bibnamefont {Lee}},\ }\href@noop {} {\bibfield
  {journal} {\bibinfo  {journal} {J. Phys. Chem. Lett.}\ }\textbf {\bibinfo
  {volume} {9}},\ \bibinfo {pages} {1734} (\bibinfo {year} {2018})}\BibitemShut
  {NoStop}%
\bibitem [{\citenamefont {Dyre}(2016)}]{dyre2016simple}%
  \BibitemOpen
  \bibfield  {author} {\bibinfo {author} {\bibfnamefont {J.~C.}\ \bibnamefont
  {Dyre}},\ }\href@noop {} {\bibfield  {journal} {\bibinfo  {journal} {J. Phys.
  Condens. Matter}\ }\textbf {\bibinfo {volume} {28}},\ \bibinfo {pages}
  {323001} (\bibinfo {year} {2016})}\BibitemShut {NoStop}%
\bibitem [{\citenamefont {Rosenfeld}(1977)}]{rosenfeld1977relation}%
  \BibitemOpen
  \bibfield  {author} {\bibinfo {author} {\bibfnamefont {Y.}~\bibnamefont
  {Rosenfeld}},\ }\href@noop {} {\bibfield  {journal} {\bibinfo  {journal}
  {Phys. Rev. A}\ }\textbf {\bibinfo {volume} {15}},\ \bibinfo {pages} {2545}
  (\bibinfo {year} {1977})}\BibitemShut {NoStop}%
\bibitem [{\citenamefont {Jakse}\ and\ \citenamefont
  {Pasturel}(2016)}]{jakse2016excess}%
  \BibitemOpen
  \bibfield  {author} {\bibinfo {author} {\bibfnamefont {N.}~\bibnamefont
  {Jakse}}\ and\ \bibinfo {author} {\bibfnamefont {A.}~\bibnamefont
  {Pasturel}},\ }\href@noop {} {\bibfield  {journal} {\bibinfo  {journal} {Sci.
  Rep.}\ }\textbf {\bibinfo {volume} {6}},\ \bibinfo {pages} {20689} (\bibinfo
  {year} {2016})}\BibitemShut {NoStop}%
\bibitem [{\citenamefont {Carnahan}\ and\ \citenamefont
  {Starling}(1969)}]{carnahan1969equation}%
  \BibitemOpen
  \bibfield  {author} {\bibinfo {author} {\bibfnamefont {N.~F.}\ \bibnamefont
  {Carnahan}}\ and\ \bibinfo {author} {\bibfnamefont {K.~E.}\ \bibnamefont
  {Starling}},\ }\href@noop {} {\bibfield  {journal} {\bibinfo  {journal} {J.
  Chem. Phys.}\ }\textbf {\bibinfo {volume} {51}},\ \bibinfo {pages} {635}
  (\bibinfo {year} {1969})}\BibitemShut {NoStop}%
\end{thebibliography}%
\end{document}